\documentclass[11pt,reqno,fleqn]{article}
\usepackage{lscape,amssymb, amsmath,verbatim,graphicx}
\usepackage{epsfig,subfigure,float}
\usepackage{morefloats}
\usepackage{enumerate}
\usepackage{booktabs}
\usepackage{amsthm}
\usepackage{color}

\definecolor{db}{rgb}{0,0,0}

\usepackage[font =footnotesize, textfont = footnotesize]{caption}

\theoremstyle{plain}
\newtheorem{theorem}{Theorem}
\newtheorem{lemma}{Lemma}

\usepackage    {amsmath}
\usepackage    {amssymb}
\usepackage    {epsf}
\usepackage    [longnamesfirst]{natbib}
\bibpunct{(}{)}{,}{a}{ }{;}
\usepackage    {amsthm}
\usepackage[T1]{fontenc}
\usepackage{lato}

\newfont{\popis}{cmcsc10}

\makeatletter
\renewcommand\@seccntformat[1]{{\csname the#1\endcsname}.~}
\makeatother
\theoremstyle{plain}

\theoremstyle{remark}

\theoremstyle{definition}

\newcommand{\carka}{\raise0.2em\hbox{,}}

\newcommand{\begeqO}{\begin{eqnarray*}}
\newcommand{\eneqO}{\end{eqnarray*}}
\newcommand{\begeq}{\begin{eqnarray}}
\newcommand{\eneq}{\end{eqnarray}}

\newcommand{\nin}{\noindent}

\newcommand{\bx}{\boldsymbol x}
\newcommand{\bbeta}{\boldsymbol \beta}

\newcommand{\bDelta}{\boldsymbol \Delta}

\newcommand{\bd}{\boldsymbol d}

\newcommand{\bC}{\boldsymbol C}

\newcommand{\what}{\widehat}

\newcommand{\oh}{1/2}
\newcommand{\ofrac}[1]{\frac{1}{#1}}

\newcommand{\R}{\mathbb{R}} 

\newcommand{\g}{\gamma}

\newcommand{\prendwol}{\hfill $\square$}

\newcommand\relphantom[1]{\mathrel{\phantom{#1}}} 
\newcommand{\mbf}[1]{\mathbf{ #1}}
\newcommand{\eps}{\varepsilon}
\newcommand{\E}{\textnormal{E}}

\newcommand{\Var}{\textnormal{Var}}
\newcommand{\Cov}{\textnormal{Cov}}
\newcommand{\bs}[1]{\boldsymbol{ #1}}

\newcommand{\oinO}{$\omega\in\Omega$ }
\newcommand{\ceil}[1]{\lceil #1\rceil}
\newcommand{\bfrac}[2]{\left(\frac{#1}{#2}\right)}

\newcommand{\as}{\quad\text{a.s.}}
\newcommand{\limm}{\lim\limits_{m\rightarrow\infty}}

\newcommand{\mtoinf}{m\rightarrow\infty}
\newcommand{\limarm}{\stackrel{(m\rightarrow\infty)}{\longrightarrow}}
\newcommand{\limP}{\stackrel{P}{\longrightarrow}}
\newcommand{\eqD}{\stackrel{\mathcal{D}}{=}}
\newcommand{\abs}[1]{\left| {#1} \right|}
\newcommand{\lr}[1]{\left({#1}\right)}

\newcommand{\op}{o_P(1)}

\newcommand{\Op}{\mathcal{O}_P(1)}
\newcommand{\Opo}[1]{\mathcal{O}_P\left(#1\right)}
\newcommand{\bigO}[1]{\mathcal{O}\left(#1\right)}

\newcommand{\page}{\textnormal{Page}}

\newcommand{\betam}{\what{\bbeta}_m}

\newcommand{\emb}{\overline{\eps}_m}

\newcommand{\km}{\frac{k}{m}}

\newcommand{\tm}{\frac{t}{m}}
\newcommand{\sm}{s/m}
\newcommand{\tmc}{\ceil{t}/m}
\newcommand{\smc}{\ceil{s}/m}

\newcommand{\kb}{k^{*}}
\newcommand{\hag}{h_{\alpha,\g}(m,k)}
\newcommand{\gtilde}{g(m,k)}

\newcommand{\gtildem}[1]{g(m,#1)}

\newcommand{\gkm}{\left(1+k/m\right)\left(k/({k+m})\right)^\gamma}

\newcommand{\gt}{\left(1+t\right)\left(t/(1+t)\right)^\gamma}

\newcommand{\gT}[1]{u({#1})}

\newcommand{\supk}{\sup\limits_{1\leq k<\infty}}
\newcommand{\supt}{\sup\limits_{0< t<\infty}}

\newcommand{\mTkmax}{\max\limits_{1\leq k\leq mT}}
\newcommand{\supmTk}{\sup\limits_{mT\leq k<\infty}}
\newcommand{\supmTt}{\sup\limits_{mT\leq t<\infty}}

\newcommand{\suptT}{\sup\limits_{0< t\leq T}}
\newcommand{\supTt}{\sup\limits_{T\leq t<\infty}}

\newcommand{\minnik}{\min\limits_{0\leq i\leq k}}

\newcommand{\maxnik}{\max\limits_{0\leq i\leq k}}

\newcommand{\maxnikabs}[1]{\max\limits_{0\leq i\leq k}\left|#1\right|}

\newcommand{\infst}{\inf\limits_{0\leq s\leq t}}

\newcommand{\supstc}{\sup\limits_{0\leq s\leq \ceil{t}}}

\newcommand{\supst}{\sup\limits_{0\leq s\leq t}}
\newcommand{\supstabs}[1]{\sup\limits_{0\leq s\leq t}\abs{#1}}

\newcommand{\We}[1]{W_1\left(#1\right)}
\newcommand{\Wz}{W_0\left(1\right)}
\newcommand{\Wd}[1]{W\left(#1\right)}

\newcommand{\Rpkm}{\maxnik\frac{\abs{W_1\lr{k/m}-W_1\lr{i/m} -((k-i)/{m})W_0(1)}}{\gkm}}
\newcommand{\Rptmc}{\ofrac{u(\tmc)}\supstc\abs{W_1\lr{\tmc}-W_1\lr{\smc} -(({\ceil{t}-\ceil{s}})/{m})W_0(1)}}
\newcommand{\Rptm}{\ofrac{u(t/m)}\supstabs{W_1\lr{t/m}-W_1\lr{s/m} -(({s-t})/{m})W_0(1)}}

\newcommand{\Rpik}{R_P(m,k)}

\newcommand{\Rpit}{R_P(t)}

\newcommand{\Rpitm}{\overline{R}_P(m,t)}

\newcommand{\Rpitmc}{\widetilde{R}_P(m,\ceil{t})}

\newcommand{\sume}[2]{\sum\limits_{#1 = m+1}^{m+#2}\eps_{#1}}
\newcommand{\sumei}[2]{\sum\limits_{#1 = m+i+1}^{m+#2}\eps_{#1}}

\newcommand{\sumeh}[2]{\sum\limits_{#1 = m+1}^{m+#2}\hat{\eps}_{#1}}
\newcommand{\sumehs}[2]{\sum\limits_{#1 = m+1}^{m+#2}\hat{\eps}^2_{#1}}
\newcommand{\sumem}{\sum\limits_{\ell = 1}^{m}\eps_{\ell}}

\newcommand{\sumemh}{\sum\limits_{\ell = 1}^{m}\hat{\eps}_{\ell}}
\newcommand{\sumemhs}{\sum\limits_{\ell = 1}^{m}\hat{\eps}^2_{\ell}}

\newcommand{\Q}{\what{Q}(m,k)}
\newcommand{\Qhm}[1]{\what{Q}(m,#1)}
\newcommand{\Qm}[1]{Q(m,#1)}

\newcommand{\Qpi}{\what{Q}_P(m,k)}
\newcommand{\Po}{\what{Q}_P^u(m,k)}
\newcommand{\Pt}{\what{Q}_P^d(m,k)}

\newcommand{\Qpiwh}{Q_P(m,k)}

\newcommand{\tp}{\tau^\page_{\alpha,\g}(m)}

\newcounter{A-index}
\newcommand{\aind}{\arabic{A-index}}
\newcommand{\aindp}{\stepcounter{A-index}\arabic{A-index}}

\begin{document}
\renewcommand{\thefootnote}{\fnsymbol{footnote}}
\begin{center}
\textbf{\textsc{{{\nin
\Huge
Page's Sequential Procedure\\[-2mm] 
for Change-Point Detection\\[5mm] 
in Time Series Regression
}
}
}
\renewcommand{\thefootnote}{\arabic{footnote}}}
\end{center}
\vspace{1cm}
  \begin{center}
\textbf{
\textsc{
{\LARGE{Stefan Fremdt}}\\
{\Large University of Cologne}
}
}
\end{center}

\begin{center}
{\normalsize 
{
{\Large Mathematical Institute}
\\
Weyertal 86--90, D--50\,931 K\" oln, Germany.\\
E-mail:} sfremdt@math.uni-koeln.de}
\end{center}

\medskip\nin

\begin{abstract}
\noindent 
In a variety of different settings cumulative sum (CUSUM) procedures have been applied for the sequential detection of structural breaks in the parameters of stochastic models. Yet their performance depends strongly on the time of change and is best under early-change scenarios. For later changes their finite sample behavior is rather questionable. We therefore propose modified CUSUM procedures for the detection of abrupt changes in the regression parameter of multiple time series regression models, that show a higher stability with respect to the time of change than ordinary CUSUM procedures. The asymptotic distributions of the test statistics and the consistency of the procedures are provided. In a simulation study it is shown that the proposed procedures behave well in finite samples. Finally the procedures are applied to a set of capital asset pricing data related to the Fama-French extension of the capital asset pricing model.

\vspace{8mm}


\vspace{2mm}

\noindent {\em Abbreviated Title:} Page's sequential procedure in time series regression

\vspace{2mm}


\end{abstract}

\allowdisplaybreaks

\section{Introduction}\label{sec1}
\textcolor{db}{
Linear regression models are among the most widely applied stochastic models. The spectrum of their application ranges from purely scientific to practical problems across all disciplines. It is therefore quite obvious how important statistical procedures are which reliably monitor the validity of such models. 
One example that shows impressively how exogenious shocks may lead to model misspecifications is the recent worldwide economical crisis and its aftermath. Here, shocks in the financial markets led to mispricing of assets and risks due to structural changes in the underlying stochastic models, with fatal consequences for the global economy. A fast detection of such misspecifications is therefore doubtlessly crucial. To achieve such a fast detection, we propose sequential tests which are designed to be less sensitive to the time of change compared to already existing cumulative sum procedures.\\
Like in many other scientific disciplines, linear regression in the pricing of assets is one of the most common approaches to explain the (linear) relationship between the model variables. E.g., such a linear relation can be used to explain the behavior of an asset price by factors that explain a major part of its variation. Examples for such an approach include the famous and still widely applied capital asset pricing model (CAPM) of \citet{Sharpe1964} and \citet{lintner1965valuation} and its extension proposed by \citet{Fama1993}. This multifactor extension, in contrast to the one-factor CAPM, uses two factors in addition to the market excess return to explain a higher proportion of the variation of the asset price. It will be investigated as a data example in Section \ref{sec4}.}\\ 
In the literature the change-point problem for linear models has been discussed extensively. While most of the contributions are made from an a-posteriori point of view (we refer to, e.g., \citet{bai1997}, \citet{perron2006} and \citet{CH1997}), recently the sequential or on-line change-point detection has received more and more attention. \citet{antoch2002} give a bibliographical overview of the field of on-line statistical process control. \textcolor{db}{A recent review on the detection of structural breaks in time series, with particular emphasis on CUSUM-type procedures, is given in \citet{AH2013}.}\\
The basis for this work is given in the articles of \citet{1996}, \citet{2004} and \citet{AHHK2006} who suggest cumulative sum (CUSUM) procedures in different stochastic models. CUSUM procedures work best for relatively early changes but show a slower reaction the later the change occurs. \citet{AHR2007} provided the asymptotic normality of the suitably normalized stopping time of the CUSUM procedure in a similar setting as will be considered in this work but only in a relatively small range after the start of the monitoring. The procedures that will be developed here found on an idea of \citet{1954} and should give a higher stability with respect to the time of change. Other approaches that tackle this task are so-called moving sum (MOSUM) procedures that were studied by, e.g., \citet{aue2008reaction} and \citet{chu1995mosum}. Their drawback is a strong dependence on the choice of the parameters, in particular the right choice of the window size by the statistician.\\
For the applicability to, e.g., financial problem settings 
we want to explicitly allow for certain dependencies, i.e.\ we will include many of the commonly applied time series models for the error terms as well as for the regressors in our setting. Other contributions assuming dependencies are given by, e.g., \citet{SS2010} who considered strongly mixing error terms in a linear model or \citet{HPS2007} who studied autoregressive time series in a closed-end setting.\\
The paper is organized as follows. In Section \ref{sec2} the linear model and the underlying assumptions are introduced. Section \ref{sec3} contains the definition of the detectors and stopping times as well as the results on the asymptotic distribution under the null hypothesis and the asymptotic consistency of the procedures. In Section \ref{sec4} we will present a simulation study and the results of an application of the procedures to the aforementioned Fama-French model. We conclude the paper with the proofs of our main results in Section \ref{sec5}.

\section{Model description and assumptions}\label{sec2}

Consider the linear model:
\begin{equation}
y_i = \mbf{x}_i^T \bbeta_i+\eps_i,\quad 1\leq i<\infty,\label{MD} 
\end{equation}
where $\mbf{x}_i$ is a $p\times 1$ random vector and $\bbeta_i\in\R^p$.\\ We assume that for the first $m$ observations the so-called ``noncontamination assumption'' (cf. \citet{1996}) holds, i.e. 
\begin{align}
\bbeta_i = \bbeta_0,\quad 1\leq i\leq m.\label{1}
\end{align}
As mentioned before, the constancy of the regression parameters $\bbeta_i$ in time should be tested which leads to the null hypothesis 
\begin{align*} H_0: \quad&\bbeta_i=\bbeta_0,\quad i=m+1, m+2,\ldots. \\
\intertext{We consider alternatives of one abrupt change in the regression parameter at an unknown change-point, i.e.}
H_A:\quad &\textnormal{there is }\kb\geq 1\textnormal{ such that }\bbeta_i = \bbeta_0,\quad m< i< m+\kb\notag \\
&\textnormal{and }\bbeta_i = \bbeta_*,\quad i = m+\kb, m+\kb+1,\ldots\quad\textnormal{with }\bDelta = \bbeta_* - \bbeta_0 \neq \mbf{0}.
\end{align*}
The detection procedures to be presented here will be defined via stopping times $\tau(m)$ (the detailed definition is postponed to Section \ref{sec3} of this article) chosen in such a way that under the null hypothesis:
\begin{equation}
\lim_{m\rightarrow\infty} P(\tau(m) < \infty)=\alpha,\quad 0<\alpha<1\label{F1}
\end{equation}
and under the alternative
\begin{equation}
\lim_{m\rightarrow\infty} P(\tau(m) < \infty)=1.\label{Cons} 
\end{equation}

We assume the following conditions on the regressors and the error terms:
\begin{align}
&\{\mbf{x}_i\}\text{ is a stationary sequence.}\label{4}\tag*{{\bf (A.1)}}\\
 &\mbf{x}_i^T = (1,x_{2i},\ldots,x_{pi}),\quad 1\leq i<\infty,\label{fc}\tag*{{\bf (A.2)}}\\
 &\{\eps_i,\; 1\leq i <\infty\}\quad\text{and}\quad\{\mbf{x}_i,\; 1\leq i< \infty\}\text{ are independent.}\label{6}\tag*{{\bf (A.3)}}\\
 &\text{There exist a }p\text{-dimensional vector }\bd = (d_1,\ldots,d_p)^T\text{ and constants }K>0,\relphantom{\nu>2\text{ such that}}\notag\\
&\nu>2\text{ such that}\notag\\
&\quad \E\left|\sum_{i=1}^k (x_{i,j}-d_j)\right|^\nu\leq K\,k^{\nu/2},\quad 1\leq j\leq p.\label{5}\tag*{{\bf (A.4)}}
\\
&\text{For every $m$ there are independent Wiener processes $\{W_{1,m}(t): t\geq 0\}$}\tag*{{\bf (A.5)}}\label{7}\\
&\text{and $\{W_{0,m}(t): t\geq 0\}$ and a constant $\sigma >0$ such that}\notag\\
&\supk \frac{1}{k^{\xi}} \left|\sume{i}{k} - \sigma W_{1,m}(k)\right| = \Op\quad(\mtoinf)\label{IP-I}\\
&\text{and}\notag\\
&\sumem  - \sigma W_{2,m}(m) = \Opo{m^\xi}\quad(\mtoinf),\label{IP2}\\
&\text{with some $\xi < \oh$.}\notag
\end{align}
The above stated assumptions on the regressors and error terms are satisfied for a variety of important stochastic models. For examples we refer to \citet{AHR2007} who showed that \ref{4} and \ref{5} are satisfied for, e.g., i.i.d. sequences, linear processes or augmented GARCH sequences. The latter were introduced by \citet{Duan1997} and include most of the conditionally heteroskedastic models used in practice. For a collection of examples belonging to this class we suggest the papers of \citet{ABH2006} and \citet{carrasco2002}. Concerning the error terms \citet{AHHK2006} provided the proof of \ref{7} again for augmented GARCH sequences under appropriate assumptions, \citet{AH2004} give further examples, besides the i.i.d.\ case, including martingale difference sequences and stationary mixing sequences.\\ 
All procedures treated in this work are based on the behavior of the residuals of the model
\begin{equation*}
\hat{\eps}_i = y_i - \mbf{x}_i^T\betam,\quad i = 1,2,\ldots, 
\end{equation*}
where $\what{\bbeta}_m$ denotes a $\sqrt{m}$-consistent estimator for $\bbeta$ from the data $(y_1,\mbf{x}_1),\ldots,(y_m,\mbf{x}_m)$.
In the sequel we will by $\hat{\sigma}_m$ denote a weakly consistent estimator for the parameter $\sigma$ from Assumption \ref{7}. The estimation of this parameter will be discussed later in detail.

\section{Sequential testing procedures and asymptotic results}\label{sec3}

Many sequential detection procedures in the literature are constructed as first passage times of a so-called detector over a certain boundary function. For example \citet{2004} proposed as a detector the (ordinary) CUSUM of the residuals, i.e.
\begin{align}
&\what Q(m,k) = \sum_{m<i\leq m+k}\hat{\eps}_i,\quad k=1,2,\ldots,\quad\text{ and }\quad\what Q(m,0) = 0,\notag\\
\intertext{and as a boundary function}
&h_{\alpha,\g}(m,k) = c\,\gtilde = c\,m^{1/2}\left(1+\frac{k}{m}\right)\left(\frac{k}{k+m}\right)^\gamma,\label{g}\\
\intertext{with} 
&0\leq\gamma<\oh
\label{9}
\end{align}
and $c = c(\alpha,\g)$ such that \eqref{F1} holds. It should be noted that in the sequel of this article we use the convention $\sum_{i = a}^b c_i = 0$ for $a>b$ and all real $c_i$. The first procedure we want to introduce goes back to an idea of \citet{1954} and we define the detector
\begin{align}
\Qpi &= \maxnik\left|\what Q(m,k) - \what Q(m,i)\right| = \max\left\{\Po,\Pt\right\},\label{QP-def}\\
\intertext{where }
\Po &= \what Q(m,k) - \minnik \what Q(m,i)\quad\text{and}\notag\\
\Pt &= \maxnik \what Q(m,i) - \what Q(m,k).\notag
\end{align}
The corresponding stopping time is then given by
\begin{align*}
 \tp = \inf\left\{k\geq 1:\Qpi>\hag\right\}
\end{align*}
where $\inf \emptyset = \infty$ and the constant $c = c(\alpha,\gamma)$ in the definition of $h_{\alpha,\g}$ can be derived from Theorem \ref{Th2.1-ts} below.
\begin{theorem}\label{Th2.1-ts}
 Assume that \eqref{1}, \ref{4} -- \ref{7} and \eqref{9} hold. Then under the null hypothesis we have, for $c\in\R$,
\begin{equation*}\lim_{m\rightarrow \infty} P\left(\frac{1}{\hat{\sigma}_m}\sup_{1\leq k<\infty}\frac{\Qpi}{\gtilde}\leq c\right)= P\left( \sup_{0< t< 1}\sup_{0\leq s\leq t} \frac{1}{t^\g}\left|W(t) - \frac{1-t}{1-s}W(s)\right|\leq c\right),
\end{equation*}
where $\{W(t):t\geq 0\}$ is a standard Wiener process.
\end{theorem}

\citet{1954} proposed a detector of the type $\Po$ for one-sided change-in-the-mean alternatives. In the case of a linear model this detector is appropriate for alternatives with $\bDelta^T \bs{d} > 0$, where the vector $\bs{d}$ was introduced in \ref{5}. The corresponding asymptotic result under the null hypothesis for these one-sided detectors is given in Theorem \ref{Th2.1}.
\begin{theorem}\label{Th2.1}
 Assume that \eqref{1}, \ref{4} -- \ref{7} and \eqref{9} hold. Then under the null hypothesis we have, for $c\in\R$,
\begin{align*}
&\lim_{m\rightarrow \infty} P\left(\frac{1}{\hat{\sigma}_m}\sup_{1\leq k<\infty}\frac{\Po}{\gtilde}\leq c\right)\\
= &\lim_{m\rightarrow \infty} P\left(\frac{1}{\hat{\sigma}_m}\sup_{1\leq k<\infty}\frac{\Pt}{\gtilde}\leq c\right)\notag\\
= & ~P\left( \sup_{0< t< 1} \ofrac{t^\g}\lr{W(t)-\inf_{0\leq s\leq t}\frac{1-t}{1-s}W(s)}\leq c\right),\notag
 \end{align*}
where $\{W(t):t\geq 0\}$ is a standard Wiener process.
 \end{theorem}
From this result again the critical value $c(\alpha,\g)$ can be derived for the two one-sided detectors. We will denote this critical value by $c_1 = c_1(\alpha,\g)$ and for the two-sided detector by $c_2=c_2(\alpha,\g)$.
Under the alternative hypothesis the detectors diverge as the following theorem shows.
\begin{theorem}\label{Th2.2}
Assume that \eqref{1}, \ref{4} -- \ref{7} and \eqref{9} hold.
\begin{enumerate}[a)]
\item Then under $H_A$ and if $\bd^T\bDelta>0$ we have
 \begin{align}
\frac{1}{\hat{\sigma}_m}\sup_{1\leq k<\infty}\frac{\Po}{g(m,k)} \limP \infty\quad\text{as $m\rightarrow \infty$} ,\notag
\end{align}
\item Then under $H_A$ and if $\bd^T \bDelta < 0$ we have
 \begin{align}
\frac{1}{\hat{\sigma}_m}\sup_{1\leq k<\infty}\frac{\Pt}{g(m,k)} \limP \infty\quad\text{as $m\rightarrow \infty$} ,\notag
\end{align}
\item Then under $H_A$ and if $\bd^T \bDelta \neq 0$ we have
\begin{align}
\frac{1}{\hat{\sigma}_m}\sup_{1\leq k<\infty}\frac{\Qpi}{g(m,k)} \limP \infty\quad\text{as $m\rightarrow \infty$.}\notag
\end{align}
\end{enumerate}
\end{theorem}
Theorem \ref{Th2.2} gives a sufficient condition that guarantees \eqref{Cons}. In Section \ref{sec4} tables with simulated critical values for selected values of $\alpha$ and $\g$ can be found for the functionals
\begin{align*}\sup_{0< t< 1}\ofrac{t^\gamma}\left(\Wd{t} - \infst \frac{1-t}{1-s}\Wd{s}\right)
\intertext{and}
\sup_{0< t< 1}\sup_{0\leq s\leq t} \frac{1}{t^\g}\left|W(t) - \frac{1-t}{1-s}W(s)\right|.
\end{align*}
\textcolor{db}{By construction, the Page CUSUM detector should be less sensitive to the time of change compared to the ordinary CUSUM detector. This stability can be achieved for other CUSUM-type detectors by applying the construction principle from above. In the sequel of this section we will illustrate this in the context of the latter linear model for the CUSUM of squared residuals. Their use is motivated by the additional assumptions on the \textcolor{db}{magnitude} of change (i.e.\ $\bd^T \bDelta \gtrless 0$ resp. $\bd^T \bDelta \neq 0$) for the above developed procedures. These guarantee their consistency but are quite restrictive.} Under additional assumptions on the error terms we can modify the presented procedures which allows to drop these assumptions on the \textcolor{db}{magnitude} of change. In this context we want to refer to the work of \citet{huskova2005} who with the same intention developed monitoring procedures based on quadratic forms of weighted cumulative sums.\\
\textcolor{db}{Aside from the linear model, CUSUMs of squared residuals are of use in other situations as well. E.g., \citet{ADFS2013} show their applicability to the detection of general parameter changes in autoregressive moving average time series where ordinary CUSUM procedures may only be used to detect changes in the mean  of such time series.}\\
We analogously define the Page detectors based on the cumulative sum of squared residuals via
\begin{align*}
\what{S}_P(m,k) &= \maxnik\left|\what{S}_R(m,k) - \what{S}_R(m,i)\right|
\intertext{and}
\what{S}_P^u(m,k) &= \maxnik \lr{\what{S}_R(m,k) - \what{S}_R(m,i)},
\intertext{where } 
\what{S}_R(m,k) &= \sumehs{i}{k} - \km \sumemhs, \quad k = 1,2,\ldots,\quad\text{ and }\quad \what S_R(m,0) = 0.
\end{align*}
\citet{AHHK2006} showed similar results on the squared prediction errors using the following additional assumptions:
\begin{align}
& \textcolor{db}{\{\eps_i\}\text{ is an orthogonal martingale difference sequence with respect to a}}\notag\\
& \textcolor{db}{\text{filtration }\{\mathcal{G}_i\}\text{ with }E \eps_i^2 = \sigma^2,\quad 0<\kappa = E \eps_i^4 < \infty\quad(i\geq 1)\tag*{{\bf (A.6)}}}\label{14}\\
& \textcolor{db}{\eta^2 = \Var(\eps_0^2) + 2\sum_{i=1}^\infty \Cov(\eps_0^2,\eps_i^2) > 0 }\tag*{{\bf (A.7)}}\label{15}\\
&
\textcolor{db}{
\text{There exist a positive definite matrix $\bC$ and a constant $\kappa>0$ such that }
}
\notag\\
&
\textcolor{db}{
\left|\ofrac{n}\sum_{i = 1}^n \bx_i\bx_i^T - \bC\right| = \bigO{n^{-\kappa}}\quad\text{a.s.}\quad (n\to\infty)
}
\tag*{{\bf (A.8)}}\label{erg}
\\
&
\textcolor{db}{\text{and under $H_A$ }\notag
}
\\
&
\textcolor{db}{
\ofrac{\nu-\kb}\sum_{i = m+\kb + 1}^{m+\nu} \bx_i\bx_i^T \to\bC\quad\text{a.s.}\quad \text{as }\min\{\nu-\kb,m\}\to\infty.\tag*{{\bf (A.9)}}
}
\label{erg-alt}
\end{align}
Furthermore, they assumed that for every $m$ there exist independent Wiener processes 
\begin{align*}
\{W_{3,m}(t): t\geq 0\}\quad\text{ and }\quad\{W_{4,m}(t): t\geq 0\},
\end{align*}
such that
\begin{align}
 \supk \frac{1}{k^\zeta}\left|\sum_{i = m+1}^{m+k}(\eps_i^2 - \sigma^2) - \eta W_{3,m}(k)\right| = \Op\quad(\mtoinf)\label{16}
 \intertext{and}
 \sum_{i = 1}^m(\eps_i^2 - \sigma^2) - \eta W_{4,m}(m) = \Opo{m^\zeta}\quad(\mtoinf)\label{17}
\end{align}
with some $\zeta < \oh$ and $\eta$ from \ref{15}.
Combining the techniques of \citet{AHHK2006} and from the proof of Theorem \ref{Th2.1-ts} it is obvious that similar asymptotic results hold for these procedures: 
\begin{theorem}\label{Th3.4}
 Assume that \eqref{1}, \ref{4} -- \ref{6}, \eqref{9}, \ref{14} -- \ref{erg} as well as \eqref{16} and \eqref{17} hold. Then under the null hypothesis we have, for a real number $c$ and a standard Wiener process $\{W(t):t\geq 0\}$,
\begin{align*}
&\textcolor{db}{\lim_{m\rightarrow \infty} P\left(\frac{1}{\eta}\sup_{1\leq k<\infty}\frac{\what S_R(m,k)}{\gtilde}\leq c\right)= P\left( \sup_{0< t< 1} \frac{W(t)}{t^\gamma}\leq c\right),}\\
&\lim_{m\rightarrow \infty} P\left(\frac{1}{\eta}\sup_{1\leq k<\infty}\frac{|\what S_R(m,k)|}{\gtilde}\leq c\right)= P\left( \sup_{0< t< 1} \frac{|W(t)|}{t^\gamma}\leq c\right)
\intertext{and}
&\lim_{m\rightarrow \infty} P\left(\frac{1}{\eta}\sup_{1\leq k<\infty}\frac{\what S_P(m,k)}{\gtilde}\leq c\right)= P\left( 
\sup_{0< t< 1}\sup_{0\leq s\leq t} \frac{1}{t^\g}\left|W(t) - \frac{1-t}{1-s}W(s)\right|
\leq c\right).
\end{align*}
\end{theorem}
The parameter $\eta$ in the statement of Theorem \ref{Th3.4} can be replaced by a weakly consistent estimator $\hat\eta_m$. \citet{AHHK2006} pointed out that the Bartlett estimator $\hat\eta_{B,m}^2$ for $\eta^2$ under the conditions of Theorem \ref{Th3.4} satisfies $\hat{\eta}^2_{B,m} \to \eta^2$ (in probability) and can therefore be applied in the general setting of this section. The same arguments hold for the estimation of $\sigma$. However, it should be noted that the quality of the estimators affects the finite sample behavior of the procedures. This will be discussed in Section \ref{sec4}.

Under the alternative hypothesis without additional assumptions on the \textcolor{db}{magnitude} of the change we again have the desired divergence.
\begin{theorem}\label{Th3.6}
Assume that \eqref{1}, \ref{4} -- \ref{6}, \eqref{9}, \ref{14} -- \ref{erg-alt} as well as \eqref{16} and \eqref{17} 
hold.
Then under $H_A$ we have
 \begin{align}
\frac{1}{\hat\eta_m}\sup_{1\leq k<\infty}\frac{|\what S_R(m,k)|}{\gtilde} \limP \infty\quad\text{as $m\rightarrow \infty$} \notag
\intertext{and}
\frac{1}{\hat\eta_m}\sup_{1\leq k<\infty}\frac{\what S_P(m,k)}{\gtilde} \limP \infty\quad\text{as $m\rightarrow \infty$} .\notag
\end{align}
\end{theorem}
Analogous results to those of Theorems \ref{Th3.4} (with the corresponding limit distributions from Theorem \ref{Th2.1}) and \ref{Th3.6} hold for the detectors $\what S_P^u$ (and $\what S_R$). However, as we will see in Section \ref{sec4}, these show a poorer finite sample behavior than the detectors $\what S_P$ and $|\what S_R|$.\\
One drawback of the detectors $\what S_P$, $|\what S_R|$, $\what S_P^u$ and $\what S_R$ is that the assumption on the existence of a constant $\sigma$ is crucial to the testing procedure. It can be seen easily that, due to its construction, the procedure is also sensitive towards changes in $\sigma$, i.e.\ in case of constant $\bbeta_i$ but a change in $\sigma (= \sigma_i)$ the testing procedure would decide that there has been a change in the $\bbeta_i$ with probability one. This sensitivity exists as well for the further introduced procedures, although in a weaker sense, i.e.\ in the derivation of the critical values which is also strongly dependent on the assumption of a constant $\sigma$.
But, since in general practitioners are concerned with the validity of their underlying model, the detection of a switch in the regime, including $\bbeta_i$ as well as $\sigma$, is of great interest to them.

\section{Simulations and an application to asset pricing data}\label{sec4}
In this section the results of a simulation study are presented that was performed to confirm the theoretical results from Section \ref{sec3}. Furthermore it should show that the proposed monitoring procedures have the desired properties. With regard to the application to the Fama-French model and its financial context the carried out simulations will focus on GARCH regressors. We will first consider the asymptotic results from Section \ref{sec3} and provide the empirical sizes under the null hypothesis. A comparison of the detection properties of the different procedures in finite samples concludes the simulation study and highlights the advantages of the newly developed sequential tests. The last part of this section will then contain the results of an application of our monitoring procedures to a data set made publically available by Kenneth R. French on his website (cf. \citet{FF-website}).\\
To establish \eqref{F1} for the suggested procedures it is necessary to determine the critical values from the definition of $h_{\alpha,\g}$ in \eqref{g} using the statements of Theorems \ref{Th2.1-ts}, \ref{Th2.1} and \ref{Th3.4}. The critical values $c_1(\g,\alpha)$ and $c_2(\g,\alpha)$ 
for the functionals
\begin{align*}\sup_{0< t< 1}&\ofrac{t^\gamma}\left(\Wd{t} - \infst \frac{1-t}{1-s}\Wd{s}\right)
\intertext{and}
\sup_{0< t< 1}&\sup_{0\leq s\leq t} \frac{1}{t^\g}\left|W(t) - \frac{1-t}{1-s}W(s)\right|,
\end{align*}
for selected values of $\alpha$ and $\g$, can be found in Table \ref{T1} and Table \ref{T2}, respectively. These were simulated with 100,000 replications of an approximation of a Wiener process generated on a grid of 100,000 points. \citet{2004} provided the simulated critical values for the functional $\sup_{0<t<1}|W(t)|/t^\g$. For $\g = 0$ we calculated these critical values numerically using the series representation
\[P\lr{\sup_{0<t<1}|W(t)|\leq c} = \frac{4}{\pi}\sum_{k = 0}^\infty \frac{(-1)^k}{2k+1}\exp\lr{-\pi^2(2k+1)^2/8c^2},\]
 from, e.g., \citet{1981}, Theorem 1.5.1, to find:
\begin{center}
\begin{tabular}{c|c|c|c|c|c}
$\alpha$ & 0.010 & 0.025 & 0.050 & 0.100 & 0.250 \\
\hline
$c(0,\alpha)$ &  2.8070 & 2.4977 & 2.2414 & 1.9600 & 1.5341
 \end{tabular}
 \end{center}
 
\begin{table}
\centering
\begin{tabular}{llllll}
\toprule
 &  \multicolumn{5}{c}{$\alpha$} \\
\cmidrule(lr){2-6}
$\gamma$ & 0.010 & 0.025 & 0.050 & 0.100 & 0.250 \\
\midrule
0.00 &  2.5955 & 2.2564 & 1.9897 & 1.6924 & 1.2474 \\
0.15 &  2.6632 & 2.3341 & 2.0757 & 1.7915 & 1.3671 \\
0.25 &  2.7372 & 2.4206 & 2.1686 & 1.8992 & 1.4887 \\
0.35 &  2.8691 & 2.5684 & 2.3273 & 2.0757 & 1.6817 \\
0.45 &  3.1712 & 2.9224 & 2.6976 & 2.4592 & 2.0932 \\
0.49 &  3.5385 & 3.2791 & 3.0640 & 2.8225 & 2.4391 \\
\bottomrule
 \end{tabular}
\caption{Critical values $c_1=c_1(\gamma,\alpha)$ simulated on a grid of 100,000 points with 100,000 replications.} \label{T1}
\bigskip
\begin{tabular}{llllll}
\toprule
 &  \multicolumn{5}{c}{$\alpha$} \\
\cmidrule(lr){2-6}
$\gamma$ & 0.010 & 0.025 & 0.050 & 0.100 & 0.250 \\
\midrule
0.00 &  2.8262 & 2.5188 & 2.2599 & 1.9914 & 1.5918 \\
0.15 &  2.8925 & 2.5925 & 2.3416 & 2.0803 & 1.6976 \\
0.25 &  2.9638 & 2.6707 & 2.4296 & 2.1758 & 1.8063 \\
0.35 &  3.0857 & 2.8041 & 2.5758 & 2.3339 & 1.9839 \\
0.45 &  3.3817 & 3.1259 & 2.9241 & 2.7002 & 2.3685 \\
0.49 &  3.7357 & 3.4903 & 3.2848 & 3.0603 & 2.7178 \\
\bottomrule
 \end{tabular}
\caption{Critical values $c_2=c_2(\gamma,\alpha)$ simulated on a grid of 100,000 points with 100,000 replications.} \label{T2}
\end{table}
\subsection{Simulation results}
\begin{table}
\begin{center}
\begin{tabular}{lrcccccc}
\toprule
&\multicolumn{1}{l}{}&\multicolumn{2}{c}{\bf $\gamma =  0 $}&\multicolumn{2}{c}{\bf $\gamma =  0.25 $}&\multicolumn{2}{c}{\bf $\gamma =  0.49 $}\\
\cmidrule(lr){3-4}\cmidrule(lr){5-6}\cmidrule(lr){7-8}
&\multicolumn{1}{c}{$m$}&\multicolumn{1}{c}{$\alpha =  0.05 $}&\multicolumn{1}{c}{$\alpha =  0.1 $}&\multicolumn{1}{c}{$\alpha =  0.05 $}&\multicolumn{1}{c}{$\alpha =  0.1 $}&\multicolumn{1}{c}{$\alpha =  0.05 $}&\multicolumn{1}{c}{$\alpha =  0.1 $}\\ \midrule
{\bf $\widehat{Q}_P$}&$100$&0.0298&0.0698&0.0390&0.0802&0.0168&0.0334\\  
&$200$&0.0294&0.0646&0.0364&0.0760&0.0194&0.0382\\
&$500$&0.0286&0.0682&0.0368&0.0812&0.0248&0.0470\\
&$1000$&0.0300&0.0704&0.0408&0.0842&0.0272&0.0554\\
\midrule
{\bf $\widehat{Q}_P^u$}&$100$&0.0354&0.0770&0.0438&0.0818&0.0166&0.0364\\
&$200$&0.0338&0.0720&0.0418&0.0806&0.0194&0.0382\\
&$500$&0.0342&0.0752&0.0430&0.0880&0.0256&0.0466\\
&$1000$&0.0350&0.0766&0.0432&0.0852&0.0256&0.0510\\
\bottomrule
\end{tabular}
\caption{Empirical sizes of the Page CUSUM procedures for 5000 replications with a monitoring horizon of $N = 5m$.}
\label{T3}
\end{center}
\end{table}
\begin{table}
\begin{center}
\begin{tabular}{lrcccccc}
\toprule
&\multicolumn{1}{l}{}&\multicolumn{2}{c}{\bf $\gamma =  0 $}&\multicolumn{2}{c}{\bf $\gamma =  0.25 $}&\multicolumn{2}{c}{\bf $\gamma =  0.49 $}\\
\cmidrule(lr){3-4}\cmidrule(lr){5-6}\cmidrule(lr){7-8}
&\multicolumn{1}{c}{$m$}&\multicolumn{1}{c}{$\alpha =  0.05 $}&\multicolumn{1}{c}{$\alpha =  0.1 $}&\multicolumn{1}{c}{$\alpha =  0.05 $}&\multicolumn{1}{c}{$\alpha =  0.1 $}&\multicolumn{1}{c}{$\alpha =  0.05 $}&\multicolumn{1}{c}{$\alpha =  0.1 $}\\
\midrule
{\bf $\widehat{S}_P$}&$100$&0.0890&0.1358&0.1072&0.1624&0.1130&0.1408\\
&$200$&0.0598&0.0976&0.0718&0.1210&0.0912&0.1190\\
&$500$&0.0416&0.0804&0.0548&0.0998&0.0826&0.1152\\
&$1000$&0.0392&0.0794&0.0520&0.0958&0.0856&0.1178\\
\midrule
{\bf $|\widehat{S}_R|$}&$100$&0.0826&0.1246&0.1014&0.1504&0.1122&0.1408\\
&$200$&0.0550&0.0888&0.0676&0.1118&0.0898&0.1176\\
&$500$&0.0396&0.0756&0.0526&0.0916&0.0784&0.1112\\
&$1000$&0.0358&0.0748&0.0488&0.0908&0.0788&0.1152\\
\midrule
{\bf $\widehat{S}_P^u$}&$100$&0.1292&0.1932&0.1570&0.2172&0.1404&0.1836\\
&$200$&0.0898&0.1516&0.1142&0.1814&0.1174&0.1560\\
&$500$&0.0676&0.1222&0.0852&0.1416&0.1124&0.1514\\
&$1000$&0.0604&0.1094&0.0760&0.1326&0.1116&0.1578\\
\midrule
{\bf $\widehat{S}_R$}&$100$&0.1196&0.1832&0.1452&0.2082&0.1410&0.1812\\
&$200$&0.0812&0.1426&0.1066&0.1702&0.1162&0.1574\\
&$500$&0.0636&0.1178&0.0792&0.1334&0.1100&0.1492\\
&$1000$&0.0566&0.1016&0.0708&0.1222&0.1074&0.1520\\
\bottomrule
\end{tabular}
\caption{Empirical sizes of the procedures based on the squared residuals for 5000 replications with a monitoring horizon of $N = 5m$.}
\label{T4}
\end{center}
\end{table}
The simulations were performed for a selection of the above mentioned models satisfying our assumptions (cf. \citet{AHR2007}, \citet{AH2004}), but since all gave similar results, we only present the results for our model \eqref{MD} with $p = 2$, $x_{2,i}$ according to a GARCH(1,1) model and independent normally distributed errors $\eps_i$ with $\sigma^2 = 0.5$ (in this specification $\sigma = \eta$ to achieve a better comparability of the procedures based on ordinary and squared residuals under the alternative).
We followed \citet{AHR2007} and chose the specification of the GARCH(1,1) model as 
\[
 x_{2,i} = d_2 + \bar{\sigma}_i z_i,\text{ with }\bar{\sigma}\text{ given as solution of }\bar{\sigma}_i^2 = \bar{\omega} + \bar{\alpha}z_{i-1}^2 + \bar{\beta}\bar{\sigma}_{i-1}^2,
\]
where $\{z_i\}$ are iid standard normally distributed and $(\bar{\omega},\bar{\alpha},\bar{\beta}) = (0.5,0.2,0.3)$. From the decomposition \eqref{exp} in the proof of Theorem \ref{Th2.2} and a similar decomposition for the procedure based on the squared residuals we find that for this model the drift in case of a change is determined by $\bd^T\bDelta$ for the ordinary residuals and for the squared residuals (asymptotically) via $\Delta_2^2+( \bd^T\bDelta)^2$. For the simulations we chose $d_2 = 1$ and used the OLSE to estimate $\bbeta_0$. Due to the uncorrelated error terms in this model, the OLSE for the parameter $\sigma$ from Assumption \ref{7}, i.e.,
$\sqrt{\hat{\sigma}_m^2} = \left(\ofrac{m-p}\sum_{i = 1}^m\lr{\hat{\eps}_i-\ofrac{m}\sum_{\ell = 1}^m\hat{\eps}_\ell}^2\right)^{1/2},$ and the corresponding estimator for $\eta$ could be utilized as well. As mentioned above, in the general setting of this paper the Bartlett estimator is a consistent estimator in the case of correlated error terms. However, simulations have shown that due to a slower convergence of the estimator, size distortions can be observed under the null hypothesis. Consequently, larger training samples are needed to achieve satisfying results.\\
The length of the training period $m$ was chosen as $m = 100, 200, 500$ and $1000$, the number of replications as 5000. For the tuning parameter $\gamma$ the values were set to $\gamma = 0.00, 0.25, 0.49$.\\
Table \ref{T3} shows the empirical sizes of the testing procedures based on the detectors $\widehat{Q}_P$ and $\widehat{Q}^u_P$ under the null hypothesis with $\bbeta_0 = (1,1)^T$ taking $N = 5m$ observations after the end of the training period. It can be seen that for all parameter combinations the sizes remain conservative for short as well as long training periods. A similar behavior was observed for the procedures based on the ordinary CUSUM and the corresponding results are therefore omitted here.\\
The conservative nature of the empirical sizes from Table \ref{T3} cannot be found for the procedures based on the squared residuals. In Table \ref{T4} the corresponding empirical sizes are displayed which show a reasonable behavior for small values of $\g$. With increasing $\g$ the size of the training period has to increase as well to find satisfactory results. This can again be explained by the estimation error for the parameter $\eta$ and the higher sensitivity of the boundary functions at the beginning of the monitoring for larger values of $\g$. For $\g$ close to $1/2$ the empirical sizes exceed the significance levels even for the larger sample sizes. This effect of a slower convergence should be taken into account by practitioners choosing the value of $\g$ and an adaptation of the procedure to include the variation of the estimator for small samples may be considered. The detectors $\what{S}_P$ and $|\what{S}_R|$ show a nicer behavior for small samples compared to $\what{S}_P^u$ and $\what{S}_R$ (which once more is due to the estimation error mentioned above). On the other hand we will see later that these procedures provide better behavior regarding 
the speed of detection.\\
To investigate the behavior of the proposed procedures under the alternative hypothesis, extensive simulations were performed for a collection of different parameter settings. We will again only give a selection of the obtained results. Since we are interested mainly in the comparison of the speed of detection of the Page CUSUM procedures with the ordinary CUSUM procedures we will comment only briefly on the power properties of the proposed procedures. The question whether a change is detected by these procedures in this open-end setting is not as interesting with regard to the comparison of ordinary and Page CUSUM. To explain this, we again refer to the construction of the procedures. The drift induced by a change is similar for both types of procedures and the boundary functions only differ by a constant. Therefore, due to the infinite monitoring horizon, the power will be similar for both types of procedures. The results of our simulations confirm this and in this matter we refer to the literature on ordinary CUSUM procedures. We will therefore continue with the comparison of the speed of detection.\\
Changes occurring at $\kb = 1,m,5m$ were considered and the monitoring was terminated at the latest after $N = \kb + 2000$ observations (which guarantees the detection of the change in all cases). The model setting under the null hypothesis described above was used and with regard to Theorems \ref{Th2.2} and \ref{Th3.6} we chose two types of changes, $\bDelta_1 = (0,0.5)^T$ and $\bDelta_2 = (-0.8,0.8)^T$, and will denote the corresponding alternative hypothesis by $H_1$ and $H_2$. With the specification of $H_1$ the above mentioned drift terms for ordinary and squared residuals are equal and a better comparability of these procedures is achieved. $H_2$ was chosen to satisfy $\bd^T\bDelta = 0$ and therefore shows that the procedures based on squared residuals perform well in this case while the procedures based on ordinary residuals are not able to detect the change.
However, the differences in the performance and applicability of the testing procedures based on ordinary residuals and those based on squared residuals should also be discussed briefly. We want to make clear that the performance strongly depends on the \textcolor{db}{magnitude} of change. For example due to their construction it can be seen from the respective drift terms that the procedures based on quadratic residuals show a slower reaction under slight changes with $\bd^T\bDelta\neq 0$ than the procedures based on ordinary residuals whereas under larger changes for the same reason the opposite is true. In addition, the influence of the parameters $\sigma$ and $\eta$ on the drift has to be taken into account. Depending on the application in practice a combination of the two procedures may be considered to balance the advantages and disadvantages of the two types of procedures.

\setlength\floatsep{0pt}
\begin{figure}
\begin{center}
\vspace*{-0.8cm}
\makebox[0cm]{
 \includegraphics[page=1, scale = 1]{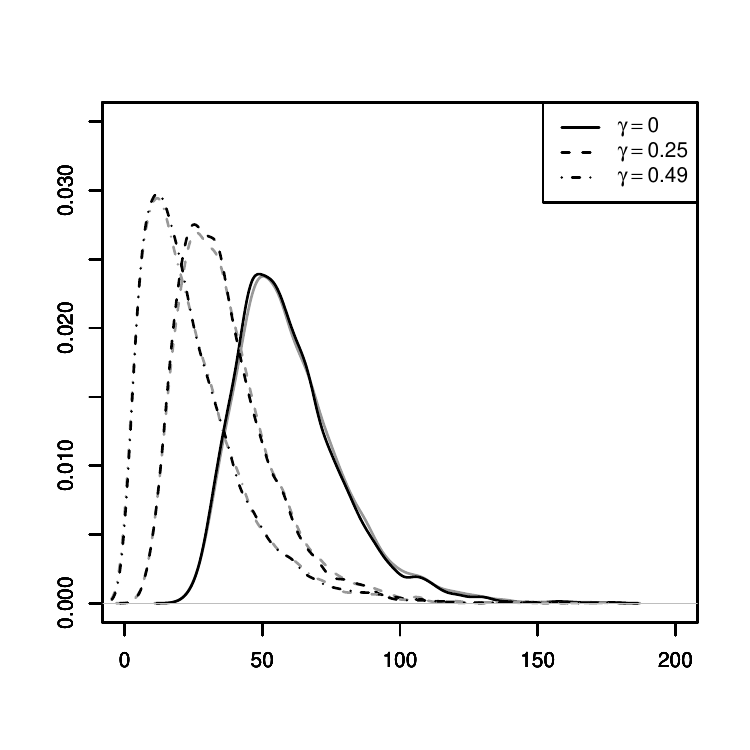}\hspace*{-5mm}
 \includegraphics[page=2, scale= 1]{plot_densities_Stat.pdf}
 }\\[-8ex]
\makebox[0cm]{
 \includegraphics[page=3, scale= 1]{plot_densities_Stat.pdf}\hspace*{-5mm}
 \includegraphics[page=4, scale= 1]{plot_densities_Stat.pdf}
 }\\[-8ex]
\makebox[0cm]{
 \includegraphics[page=5, scale= 1]{plot_densities_Stat.pdf}\hspace*{-5mm}
 \includegraphics[page=6, scale= 1]{plot_densities_Stat.pdf} 
 }\\[-4ex]
 \caption{
 Estimated density plots for the delay times under $H_1$ for $m = 100$ and $\g = 0.00,0.25,0.49$. Black lines represent Page CUSUM procedures, gray lines ordinary CUSUM procedures. The left column shows the densities of $\what{Q}_P$ and $|\what{Q}|$, the right column of $\what{S}_P$ and $|\what{S}_R|$. The rows from top to bottom represent early ($\kb = 1$), intermediate ($\kb = m$) and late ($\kb = 5m$) changes.
 }
 \label{fig:1-1}
\end{center}
\end{figure}

\begin{figure}
\begin{center}
\vspace*{-0.8cm}
\makebox[0cm]{
 \includegraphics[page=7, scale = 1]{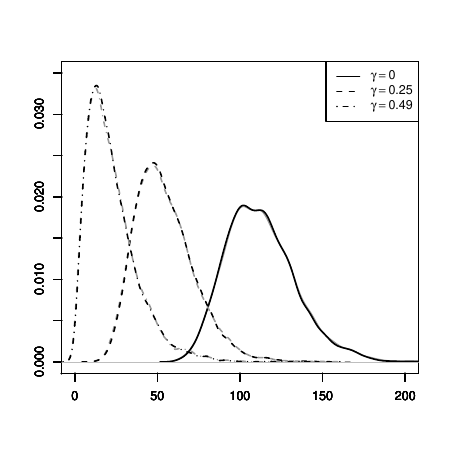}\hspace*{-5mm}
 \includegraphics[page=8, scale= 1]{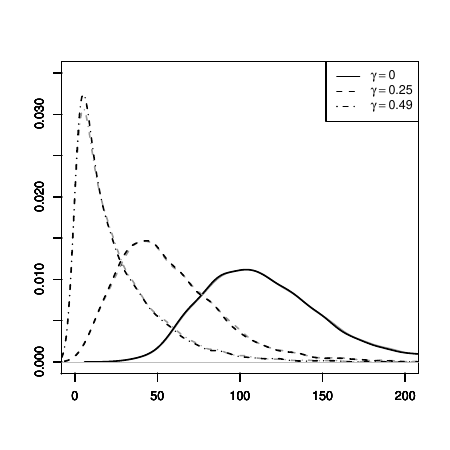}
 }\\[-8ex]
\makebox[0cm]{
 \includegraphics[page=9, scale = 1]{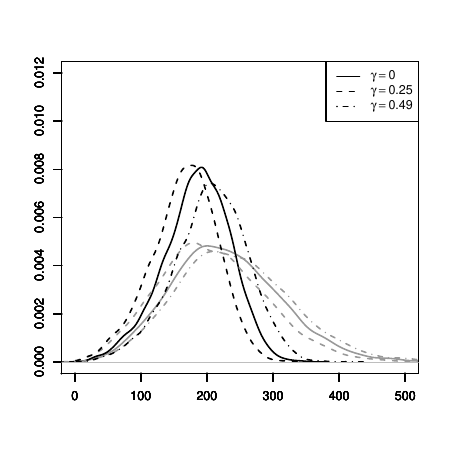}\hspace*{-5mm}
 \includegraphics[page=10, scale= 1]{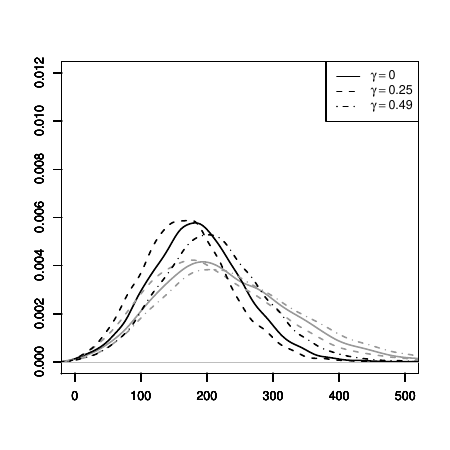}
  }\\[-8ex]
\makebox[0cm]{
\includegraphics[page=11, scale = 1]{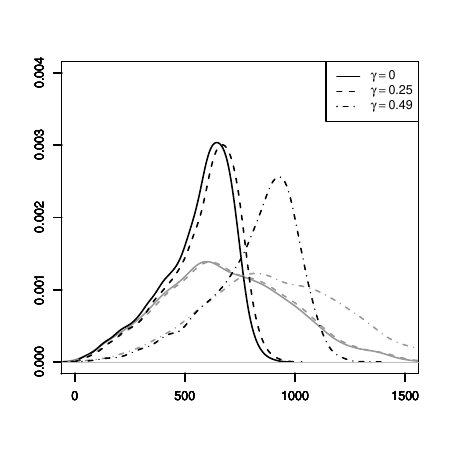}\hspace*{-5mm}
 \includegraphics[page=12, scale= 1]{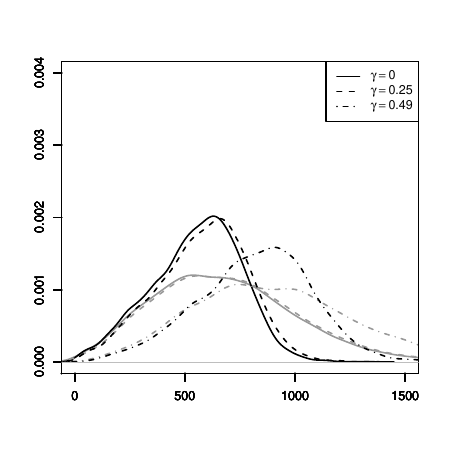}
 }\\[-4ex]
\caption{Estimated density plots for the delay times under $H_1$ for $m = 1000$ and $\gamma = 0.00,0.25,0.49$. Black lines represent Page CUSUM procedures, gray lines ordinary CUSUM procedures. The left column shows the densities of $\what{Q}_P$ and $|\what{Q}|$, the right column of $\what{S}_P$ and $|\what{S}_R|$. The rows from top to bottom represent early ($\kb = 1$), intermediate ($\kb = m$) and late ($\kb = 5m$) changes.}
 \label{fig:1-2}
\end{center}
\end{figure}

\begin{figure}
\begin{center}
\makebox[0cm]{
 \includegraphics[page=1, scale = 1]{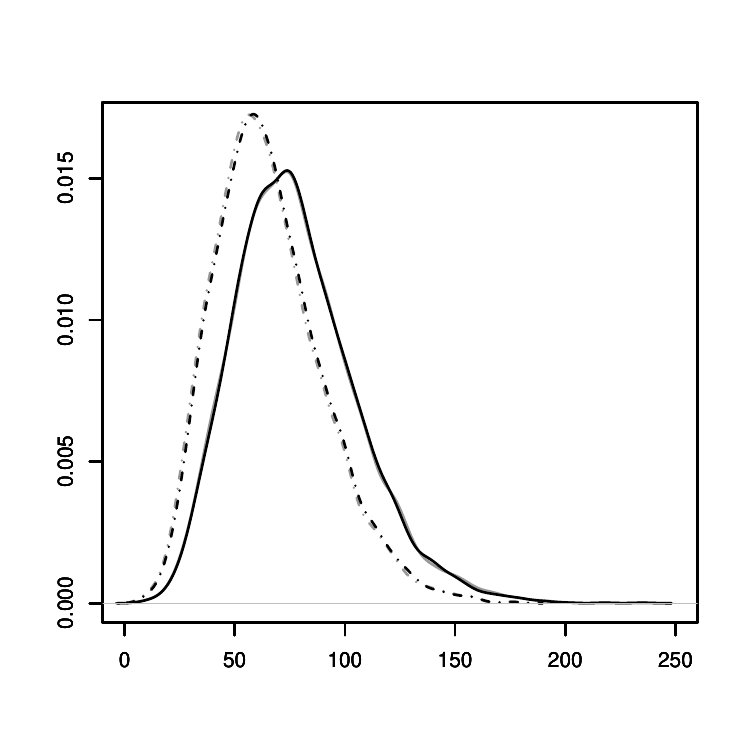}
  }\\[-8ex]
\makebox[0cm]{
 \includegraphics[page=2, scale = 1]{plot_densities_H2_Stat.pdf}
  }\\[-8ex]
\makebox[0cm]{
 \includegraphics[page=3, scale = 1]{plot_densities_H2_Stat.pdf}
 }\\[-4ex]
 \caption{Estimated density plots for the delay times under $H_2$ for $m = 1000$ and $\gamma = 0.00$ for the procedures based on squared residuals. Black lines represent Page CUSUM procedures, gray lines ordinary CUSUM procedures. Solid lines correspond to the procedures $\what{S}_P$ and $|\what{S}_R|$, dashed lines correspond to $\what{S}_P^u$ and $\what{S}_R$. The rows from top to bottom again represent early ($\kb = 1$), intermediate ($\kb = m$) and late ($\kb = 5m$) changes.
 }
\label{fig:1-3}
\end{center}
\end{figure}

We now want to illustrate that the procedures based on the Page CUSUM show a higher stability regarding the time of change than those based on the ordinary CUSUM. Additionally, the influence of the tuning parameter $\g$ on the speed of detection should be examined. Figures \ref{fig:1-1} and \ref{fig:1-2} show density estimations of the delay times (excluding false alarms) under the alternative $H_1$ for $\alpha = 0.1$. In Figure \ref{fig:1-1} a training period of length $m = 100$ was used, in Figure \ref{fig:1-2} the length was set to $m = 1000$. The rows correspond from top to bottom to very early ($\kb = 1$), intermediate ($\kb = m$) and late ($\kb = 5m$) changes. The left columns show the density estimates for $\what{Q}_P$ (black) and $|\what{Q}|$ (gray), the right columns show the estimates for $\what{S}_P$ (black) and $|\what{S}_R|$ (gray), in both columns for the different values of $\g$. \textcolor{db}{Tables containing the five number summaries of the data used for the density estimation can be found in the appendix (cf. Tables \ref{T6} -- \ref{T9}).}
\\
The density estimates show clearly that for a change immediately after the end of the training period, as could be expected, there is only a slight difference between the procedures based on Page's CUSUM and those based on the ordinary CUSUM. In this case, a choice of $\g$ close to $1/2$ gives the best results. For intermediate changes it is obvious that the Page CUSUM procedures show a better behavior than the ordinary CUSUM procedures for both ordinary and squared residuals.  This effect gets stronger the later the change occurs as can be seen in the bottom rows. For intermediate changes a choice of $\g = 0.25$ gave the best results, for late changes $\g = 0$ is the appropriate choice. This observation which reflects the role of the parameter $\g$ has already been discussed in, e.g., \citet{2004}.\\
As mentioned before, the procedures based on ordinary residuals are not applicable under the alternative $H_2$. We will therefore only present the density estimates for the procedures based on squared residuals which can be found in Figure \ref{fig:1-3}. The obtained results are similar to the results under $H_1$ regarding the comparison of Page and ordinary cumulative sums for all sample sizes. Therefore we only present these for $m = 1000$ and $\g = 0$ (where a reasonable behavior under the null hypothesis for all detectors was observed). The density estimates show that the detectors $\what{S}_R$ and $\what{S}^u_P$ detect changes faster than $|\what{S}_R|$ and $\what{S}_P$. However, due to the slower convergence to the asymptotic distribution under the null hypothesis (cf. Table \ref{T4}), their application on the basis of smaller training periods is not recommended. \\
As a conclusion of this small simulation study, we find that the proposed procedures in early-change scenarios show a similar behavior to ordinary CUSUM procedures. Yet, their advantage lies in the behavior in scenarios that include a later change. In this case the Page CUSUM procedures detect changes faster and therefore overall show a higher stability regarding the time of change. The procedures based on squared residuals need stronger moment assumptions but they work in contrast to the procedures based on ordinary residuals even under orthogonal changes. The Page CUSUM shows for these a similar behavior and can therefore be recommended. Nevertheless the procedures based on ordinary residuals in general detect small, non-orthogonal changes faster and can therefore still be of great use in practice.

\subsection{Data application: The Fama-French asset pricing model}
\begin{figure}
\begin{center}
\makebox[0cm]{
\includegraphics[scale = 0.7,page = 1]{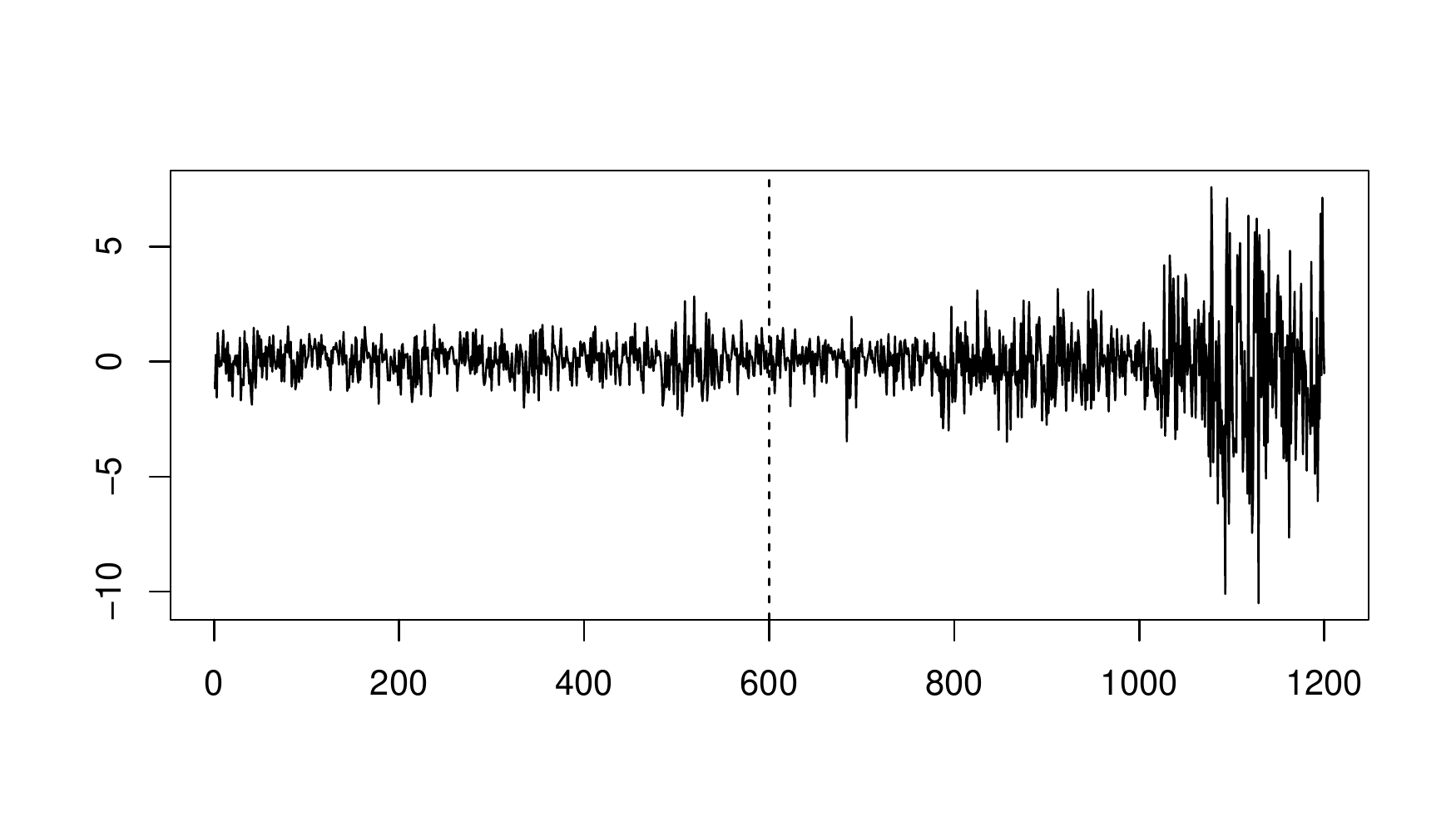}
  }\\[-8ex]
\makebox[0cm]{
\includegraphics[scale = 0.7,page = 2]{Page_FF_ts_incl-endtp-5-16.pdf}
}
\end{center}
\caption{Time series plot of the excess returns of Portfolio 1 (upper panel) and Portfolio 2 (lower panel) for the time period June 9, 2004 to March 16, 2009. The end of the training period is indicated by a dotted vertical line.}
\label{ts-pf-1-2}
\end{figure}

\begin{figure}
\begin{center}
\makebox[0cm]{
\includegraphics[scale = 0.7, page = 3]{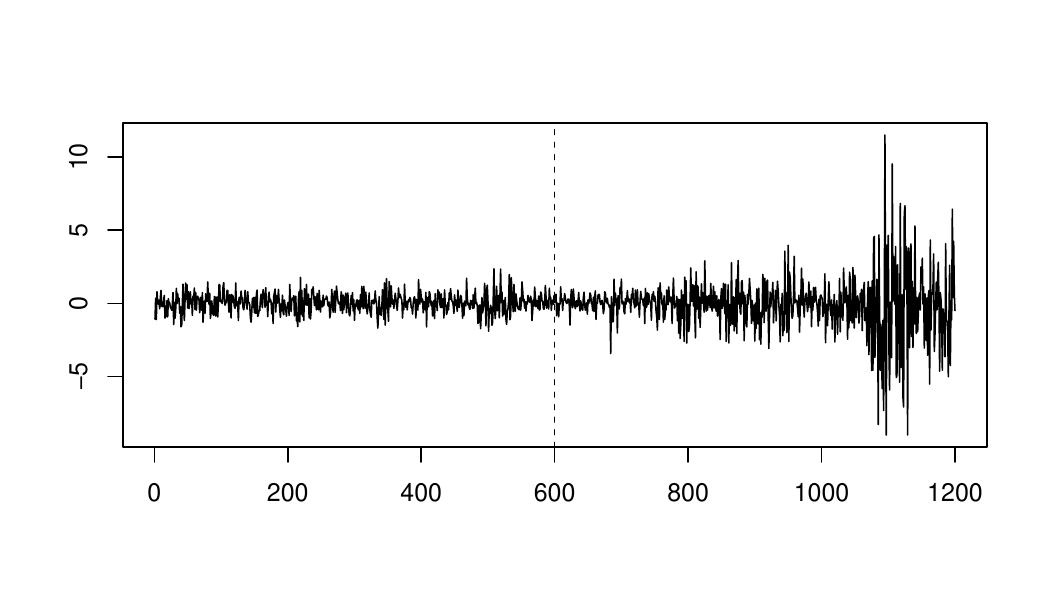}
  }\\[-8ex]
\makebox[0cm]{
\includegraphics[scale = 0.7, page = 4]{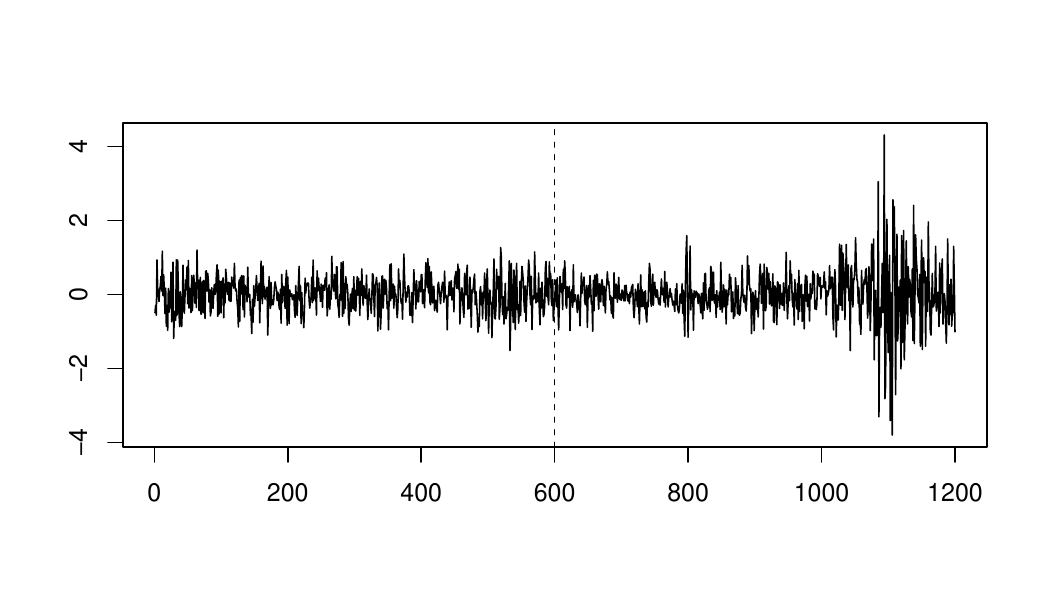}
  }\\[-8ex]
\makebox[0cm]{
\includegraphics[scale = 0.7, page = 5]{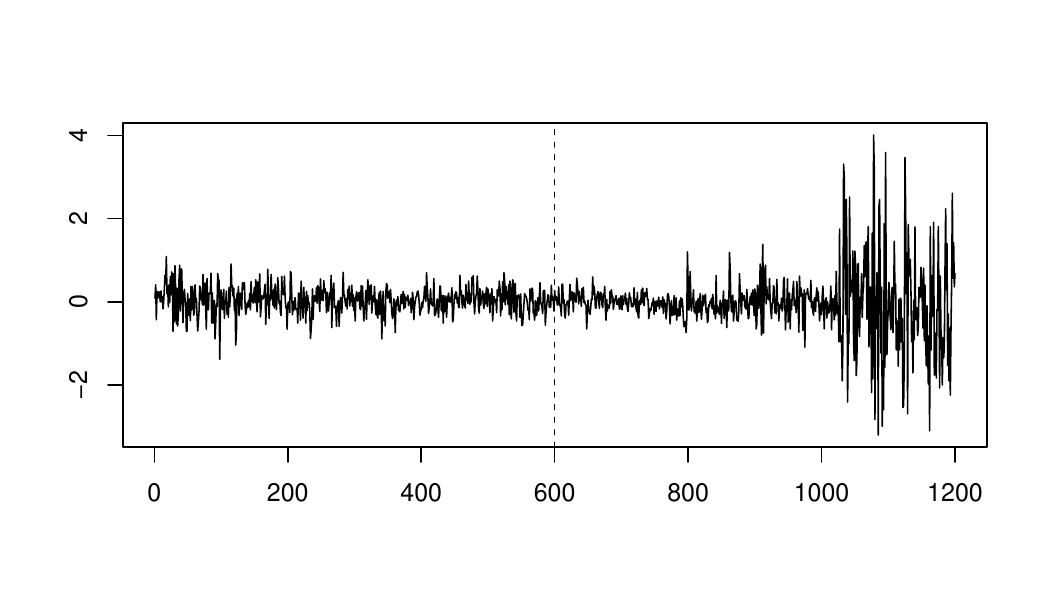}
}
\end{center}
\caption{Time series plot of the, market excess return (upper panel), size factor (middle panel) and book-to-market factor (lower panel) for the time period June 9, 2004 to March 16, 2009. The end of the training period is indicated by a dotted vertical line.}
\label{ts}
\end{figure}

\begin{figure}
\begin{center}
\makebox[0cm]{
\includegraphics[scale = 0.7,page = 2]{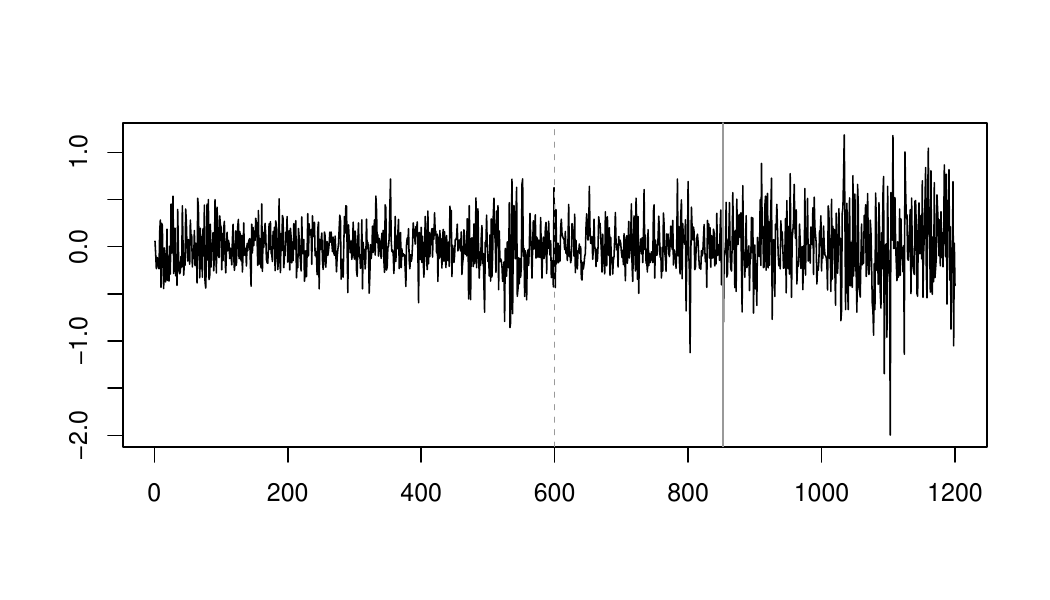}
}
\end{center}
\caption{Time series plot of the model residuals of Portfolio 2 for the time period June 9, 2004 to March 16, 2009. The end of the training period is indicated by a gray dotted vertical line, the estimated change-point by a gray solid vertical line.}
\label{ts-residuals-pf-2}
\end{figure}

In this subsection, we first want to briefly describe the asset pricing model of \citet{Fama1993}. This model tries to explain a higher proportion of the variation in the prices of asset portfolios by introducing additional factors to the capital asset pricing model of \citet{Sharpe1964} and \citet{lintner1965valuation}. \textcolor{db}{For illustration purposes, }we will then apply the monitoring procedures introduced in Section \ref{sec3} to a data set consisting of daily data for \textcolor{db}{2} asset portfolios considered by \citet{Fama1993} and the corresponding factors in the economic crisis of the years 2007 and 2008. 
\textcolor{db}{The economic context of the latter asset pricing model may be found in \citet{FF1996}, \citet{KSS1995} and \citet{MacKinlay1995}.}
For further discussion of the testing of asset pricing models for the constancy of their parameters, we want to refer, e.g., to \citet{GG1998} or \citet{AHHHS2010}.\\
\citet{Fama1993} investigated the influence of risk factors besides the market excess return on an empirical basis to explain the cross-section of average returns. As a consequence, they formulated the three-factor model for the excess return of a portfolio $i$ via
\begin{equation}
 R_i - R_f = \alpha_i + b_i(R_M - R_f) +s_i \text{SMB} +h_i \text{HML} + \eps_i,\label{FF-Model}
\end{equation}
where $R_f$ is the one month Treasury bill rate, $R_M$ is the return on the market (calculated as the value-weight return on all NYSE, AMEX and NASDAQ stocks), SMB and HML are the so called size and book-to-market factors. For a complete description of the derivation of these factors and how they are calculated we refer to \citet{Fama1993}, \citet{FF1996} and the website of Kenneth R. French (cf. \citet{FF-website}) where the underlying data set can also be found. \textcolor{db}{For our analysis we use the data of the time period June 9, 2004, to March 16, 2009 (1200 observations). As responses of this regression model we will exemplarily consider two portfolios out of a set of 25 portfolios} formed according to a categorization by size and book-to-market. For the construction of these portfolios we again refer to \citet{FF1996}. \citet{Fama1993} claim that the excess returns of these portfolios over the market are well explained by \eqref{FF-Model}. For our concerns the categorization underlying the construction is not of importance, we will consequently denote the portfolios by \textcolor{db}{Portfolio 1 and Portfolio 2. In Figure \ref{ts-pf-1-2} the time series plots of the responses $R_1 - R_f$  and $R_2 - R_f$ can be seen. Figure \ref{ts} contains the corresponding plots for the regressors $R_M-R_f$, $SMB$ and $HML$, showing obviously conditionally heteroskedastic patterns. The stopping times of detectors $\what{Q}_P$, $|\what{Q}|$, $\what{S}_P$,  $|\what{S}_R|$, $\what{S}_P^u$ and $\what{S}_R$ are displayed in Table \ref{T12}. For the tests we used $\alpha = 0.1$ and $\g = 0.25$. The length of the training period was set to $m = 600$ (i.e. until October 24, 2006, which is a relatively stable period at the markets).\\ 
For Portfolio 1 all procedures detect a change in the regression parameter. The stopping times of all detectors lie between August and December 2007 and thus at the beginning of the crisis. As already observed in the simulations, the Page CUSUM detectors perform superior to their ordinary CUSUM counterparts for all types of procedures. 
In Portfolio 2 we find a different scenario. While the procedures based on squared residuals again detect a change in the model, the procedures $\what{Q}_P$ and $|\what{Q}|$ do not react. With respect to the theory developed in the previous sections, this suggests either an orthogonal change or a change in the variance of the residuals. To further investigate this, we consider the corresponding a-posteriori CUSUM tests for the combined training and monitoring period to estimate the time of change. Table \ref{T13} shows the segment estimates for the two portfolios. For Portfolio 1 the estimates support the result of the test and show a considerable difference. For Portfolio 2 the difference of the estimates is rather small, which points to a change in the variance. This is underlined by the residual time series plot in Figure \ref{ts-residuals-pf-2} where an elevated variation can be seen in the second segment.\\
As a conclusion we find that these real life data confirm our observations from the simulation study. The Page CUSUM procedures show a superior behavior to the ordinary CUSUM procedures. With the findings of the simulation study we therefore in general suggest the use of the Page procedures.}
\begin{table}
\begin{center}
\vspace{3mm}
\begin{tabular}{lcccccc}
\toprule
Portf. No.&$\what{Q}_P$ & $|\what{Q}|$ & $\what{S}_P$ &  $|\what{S}_R|$ & $\what{S}_P^u$ & $\what{S}_R$\\
\cmidrule(lr){1-7}
~1&30/10/07&03/12/07&06/08/07&07/08/07&02/08/07&06/08/07\\
& 255 & 278 & 195 & 196 & 193 & 195\\
\cmidrule(lr){1-7}
~2&16/03/09&16/03/09&02/01/08&01/02/08&06/12/07&23/01/08\\
& 600  & 600  & 298  & 319  & 281  & 312\\
\bottomrule
\end{tabular}
\caption{Stopping times of the procedures for the Fama-French Portfolios (Dates given as dd/mm/yy in upper rows and values of the stopping times in lower rows).}
\label{T12}
\end{center}
\end{table}
\begin{table}
\begin{center}
\vspace{3mm}
\begin{tabular}{lccccc}
\toprule
Portf. No.& Period & $\alpha_i$ & $b_i$ & $s_i$ &  $h_i$ \\
\cmidrule(lr){1-6}
~1&09/06/04 -- 24/07/07& 0.0213 & 0.7637 & 0.8191 & 0.3707\\
& 24/07/07 -- 16/03/09 & -0.0753 & 0.9125 & 1.0623 & 0.5622\\
\cmidrule(lr){1-6}
~2&09/06/04 -- 25/10/07& 0.0090 & 0.9982 & 0.3032 & -0.3709\\
& 25/10/07 -- 16/03/09 & 0.0168 & 0.9712 & 0.2552 & -0.1620\\
\bottomrule
\end{tabular}
\caption{Estimated regression parameters for the two Fama-French portfolios in the segments derived from the change-point estimation (Dates given as dd/mm/yy).}
\label{T13}
\end{center}
\end{table}
\section{Proofs}\label{sec5}
 \subsection{Proof of Theorem \ref{Th2.1-ts}}
 Because of the similarity of the arguments in the proofs of Theorems \ref{Th2.1-ts} and \ref{Th2.1} we only provide the proof of Theorem \ref{Th2.1-ts}. The proof is based on a stepwise approximation of the detector $\what{Q}_P(m,k)$ from  \eqref{QP-def} via 
\begin{align}
\Qpiwh &= \maxnik\left|Q(m,k) - Q(m,i)\right|, \text{ where }\label{PQ}\\
Q(m,k) &= \sum_{m<i\leq m+k}{\eps}_i - k\emb,\quad k=1,2,\ldots,\quad\text{ and }\quad\emb = \frac 1m \sumem,\notag
\end{align}
in the first step and for every $m$ via the following functional of independent Wiener processes $\{W_{1,m}(t):t\geq 0\}$ and $\{W_{0,m}(t):t\geq 0\}$ in the second step:
\begin{equation}
W_P(m,k)= \maxnik\left|W_{1,m}(k) - W_{1,m}(i) - \frac{k-i}{m}W_{0,m}(m)\right|.\label{PW} 
\end{equation}
If not stated otherwise the asymptotics in the proofs are always assuming $m\to\infty$.
\begin{lemma}\label{L1}
If the conditions of Theorem \ref{Th2.1-ts} are satisfied then
\begin{equation*}
 \supk\ofrac{\gtilde}\left|\Qpi - \Qpiwh\right|=\op,
\end{equation*}
where $Q_P$ was defined in \eqref{PQ}.
\end{lemma}
\textbf{Proof:~}
We have
\begin{align*}
 &\abs{\maxnikabs{\Q - \Qhm{i}} - \maxnikabs{\Qm{k} - \Qm{i}}}\\
 \leq & \abs{\Q - \Qm{k}} + \maxnikabs{\Qhm{i} - \Qm{i}}.
\end{align*}
Now because $\gtilde$ increases monotonically in $k$
\begin{align*}
&\supk\frac{1}{\gtilde}\left|\Qpi - \Qpiwh\right|\\
\leq& \supk \ofrac{\gtilde}\left|\Q - \Qm{k}\right| + \supk\maxnik\ofrac{\gtildem{i}}\abs{\Qhm{i} - \Qm{i}}\\
=& 2 \supk \ofrac{\gtilde}\left|\Q-\Qm{k}\right|.
\end{align*}
It is therefore sufficient to show
\begin{equation*}
 \supk \ofrac{\gtilde}\left|\Q-\Qm{k}\right| =\op.
\end{equation*}
Using the identities
\begin{align}
 \Q &= \sume{i}{k} - \sum_{i=m+1}^{m+k}\mbf{x}_i^T(\betam - \bbeta_0)\notag\\
 \intertext{and}
 0=\sumemh &= \sumem - \sum_{i=1}^{m}\mbf{x}_i^T(\betam - \bbeta_0),\label{L1-P1a}
 \end{align}
with $\bd$ from \ref{5} we get 
\begin{align*}
\left|\Q - \Qm{k}\right|  =& \left|\lr{\frac km \sum_{i=1}^m (\mbf{x}_i-\bd)^T - \sum_{i=m+1}^{m+k} (\mbf{x}_i-\bd)^T}(\betam - \bbeta_0)\right|.
\end{align*}
In \eqref{L1-P1a} the first equality follows from the definition of $\hat{\eps}_i$ and \ref{fc}.\\
First the term $\sum_{i=1}^m (\mbf{x}_i-\bd)$ is considered. By Markov's inequality and \ref{5} it is clear that we can find \textcolor{db}{$1/2\leq\rho<1$} such that
\begin{equation}
 \left|\sum_{i=1}^m (\mbf{x}_i-\bd)\right| = \Opo{m^{\rho}}.\label{L1-P1}
\end{equation}
For the term $\sum_{i=m+1}^{m+k} (\mbf{x}_i-\bd)$ the same arguments used to show \eqref{L1-P1} apply. Then the Borel-Cantelli Lemma combined with the stationarity of the regressors yield that there exists \textcolor{db}{$1/2>\delta >0$} such that 
\begin{equation}
\left|\sum_{i=m+1}^{m+k} (\mbf{x}_i-\bd)\right| = \bigO{k^{1-\delta}} \text{ a.s., as }k\to\infty,\text{ uniformly in }m.\label{L1-P2}
\end{equation}
The $\sqrt{m}$-consistency of $\betam$ together with
\begin{equation*}
 \limm\supk \frac{km^{\rho-1} + k^{1-\delta}}{\sqrt{m}\gtilde}= 0
\end{equation*}
as well as \eqref{L1-P1} and \eqref{L1-P2} conclude the proof of Lemma \ref{L1}.\prendwol
\begin{lemma}\label{L2}
 If the conditions of Theorem \ref{Th2.1-ts} are satisfied then for each $m$ there are two independent Wiener processes $\{W_{1,m}(t):t\geq 0\}$, $\{W_{0,m}(t):t\geq 0\}$ such that
 \begin{align*}
 \supk \frac{1}{\gtilde}\left|Q_P(m,k) -  \sigma W_P(m,k)\right|=\op,
\end{align*}
where $Q_P$ and $W_P$ are as defined in \eqref{PQ} and \eqref{PW}, respectively.
\end{lemma}
\textbf{Proof:}
By similar estimations as in the proof of Lemma \ref{L1}
\begin{align*}
& \abs{Q_P(m,k) - \sigma W_P(m,k)}\\
\leq & \maxnikabs{\abs{\sumei{j}{k} - (k-i)\emb} - \sigma \abs{W_{1,m}(k) - W_{1,m}(i) - \frac{k-i}{m}W_{0,m}(m)}}\\
\leq & \abs{\sume{j}{k} - \sigma W_{1,m}(k)} + \maxnikabs{\sume{j}{i} - \sigma W_{1,m}(i)} + \frac{k}{m}\abs{\sumem - \sigma W_{0,m}(m)}
\end{align*}
and hence with assumption \ref{7}
\begin{align*}
&\supk \ofrac{\gtilde}\abs{Q_P(m,k) - \sigma W_P(m,k)}\\
\leq&\supk \ofrac{\gtilde}\abs{\sume{j}{k} - \sigma W_{1,m}(k)} \\
&+ \supk \ofrac{\gtilde}\maxnikabs{\sume{j}{i} - \sigma W_{1,m}(i)}\\
&+\supk \ofrac{\gtilde} \frac{k}{m}\abs{\sumem - \sigma W_{0,m}(m)}\\
=&~\Op\supk\frac{k^\xi}{\gtilde} + \Op\supk \frac{km^{\xi-1}}{\gtilde}\\
=&~\op,
\end{align*}
where the last equality was shown in the proof of Lemma 3 of \citet{AHHK2006}.\prendwol
\subsubsection*{Proof of Theorem \ref{Th2.1-ts}}\label{PTh2.1}
The distribution of $\{(W_{1,m}(t),W_{0,m}(t)):t\geq 0\}$ does not depend on $m$ and therefore the index can be omitted, i.e.\ we write $\{(W_1(t),W_0(t)): t\geq 0\}$ instead. 
Due to the scaling property of the Wiener process we have
\begin{align}
 \supk\frac{W_P(m,k)}{\gtilde}\eqD & \supk \Rpkm.\notag\label{pr-th21-ts-1}
\end{align}
Define
\begin{align*}
\Rpik &= \Rpkm\notag
\intertext{and together with $u(t) =\gt$ the following functionals of the Wiener processes $W_0$ and $W_1$:}
R_P(t) &= \ofrac{u(t)}\supstabs{W_1\lr{t}-W_1\lr{s} -(t-s)W_0(1)},\\
\Rpitm &= \Rptm, \quad\\ 
\Rpitmc &= \Rptmc.
\end{align*}
Note that
\begin{equation*}
\supk \Rpik = \supt \Rpitmc. 
\end{equation*}
The next step is to show:
\begin{equation}
\supk \Rpik \limarm \supt \Rpit\as.\label{pr-th21-ts-2}
\end{equation}
We divide the proof of \eqref{pr-th21-ts-2} into two steps and show
\begin{enumerate}[(i)]
\item For any $T>0$:
\begin{equation*}
\mTkmax\Rpik \limarm \suptT\Rpit\as
\end{equation*}
and
\item For almost every $\omega\in \Omega$ there exists a positive integer $T=T(\omega)$ such that
\begin{equation*}\supmTk\Rpik\limarm\supTt\Rpit.
\end{equation*}
\end{enumerate}
The first claim is immediate from the a.s.\ continuity of $R_P(t)$ on $[0,T]$ (with $R_P(0) = 0$). 
For the second claim we get for any $T>0$
\begin{align}
 &\supmTt\left|\Rpitmc-\Rpitm\right|\notag\\
 \leq&\relphantom{+}\supmTt\left|\frac{W_1\lr{\tmc}}{u(\tmc)} - \frac{W_1\lr{t/m}}{u(t/m)}\right|+\supmTt\supst \left|\frac{W_1\lr{\smc}}{u(\tmc)} - \frac{W_1\lr{\sm}}{u(t/m)}\right|\notag\\
 &+ \supmTt\supst\left|\frac{t-s}{u(t/m)}-\frac{\ceil{t}-\ceil{s}}{u(\tmc)}\right|\frac{\left|\Wz\right|}{m}\notag\\
 \leq& \,2\supmTt\supst \left|\frac{W_1\lr{\smc}}{u(\tmc)} - \frac{W_1\lr{\sm}}{u(t/m)}\right|\notag\\
 &+ \supmTt\supst\left|\frac{t-s}{u(t/m)}-\frac{\ceil{t}-\ceil{s}}{u(\tmc)}\right|\frac{\left|\Wz\right|}{m} \notag\\
 =& \,2A_{\aindp} + A_{\aindp}.\notag
\end{align}
For $A_1$ we have for any $T>0$
\begin{align*}
 &\supmTt \supst \left|\frac{\We{\smc}-\We{\sm}}{\gT{\tmc}}-\frac{\We{\sm}}{\gT{t/m}}+ \frac{\We{\sm}}{\gT{\tmc}}\right|\\
 \leq &\relphantom{+} \supmTt \supst \left|\frac{\We{\smc}-\We{\sm}}{\gT{\tmc}}\right|
 \\
 &+ \supmTt \supst \left|\We{\sm}\right|\left|\frac{1}{\gT{\tmc}}-\frac{1}{\gT{t/m}}\right|
 \\
 = & \,\supmTt \supst A_{\aindp}(t,s) + \supmTt \supst A_{\aindp}(t,s).
\end{align*}
By Theorem 1.2.1 of \citet{1981}
for all $\eps>0$ there exists a $T = T(\omega)>0$ independent of $m$ such that
\begin{equation*}\sup_{mT\leq t<\infty}\sup_{0\leq s\leq t}\sup_{0\leq r\leq 1}\frac{\left|W_1\left((s+r)/m\right) - W_1\left(\sm\right)\right|}{\gT{t/m}} <\eps\as.
\end{equation*}
Consequently for almost every \oinO there exists a $T_1 = T_1(\omega)>0$ with
\begin{align}
 &\sup_{mT_1\leq t<\infty}\supst A_3(t,s)\notag\\
 \leq & \sup_{mT_1\leq t<\infty}\supst\frac{\left|\We{\smc}-\We{\sm}\right|}{\gT{t/m}}\notag\\
 \leq &\sup_{mT_1\leq t<\infty}\sup_{0\leq s\leq t}\sup_{0\leq r\leq 1}\frac{\left|W_1\left(\frac{s+r}{m}\right) - W_1\left(\sm\right)\right|}{\gT{t/m}} <\frac{\eps}{8}.\notag
\end{align}
For $A_{\aind}$ with Theorem 1.3.1* of \citet{1981} we get similarly that for almost every \oinO there exists $T_2 = T_2(\omega)>0$ (again independent of $m$) such that
\begin{align}
\sup_{mT_2\leq t<\infty} \supst A_{\aind} \leq\sup_{mT_2\leq t<\infty} \supst \left|\frac{\We{\sm}}{\gT{t/m}}\right|<\frac{\eps}{8}.\notag
\end{align}
For $A_2$ we find that for any $T>0$:
\begin{align*}
 &\supmTt \supst \left|\frac{\ceil{t} - \ceil{s}}{\gT{\tmc}}-\frac{t - s}{\gT{t/m}}\right|\frac{\left|\Wz\right|}{m}\\
\leq &\supmTt \supst \left|\frac{\ceil{t} -t - (\ceil{s}-s)}{\gT{\tmc}}+\lr{t-s}\lr{\ofrac{\gT{\tmc}}-\ofrac{\gT{t/m}}}\right|\frac{\left|\Wz\right|}{m}\\
\leq&\supmTt \ofrac{m}\frac{\left|\Wz\right|}{\gT{\tmc}}+\supmTt \tm\left|\frac{1}{\gT{t/m}}-\frac{1}{\gT{\tmc}}\right|\left|\Wz\right|\\
=&\, A_{\aindp}.
\end{align*}
From \eqref{9} and because 
\begin{equation}
\left(m\,\gT{\tmc}\right)^{-1}\leq \left(m\,\gT{t/m}\right)^{-1},\label{pr-th21-ts-3}
\end{equation}
we have
\begin{align}
\left| \frac{t/m}{\gT{t/m}} - \frac{t/m}{\gT{\tmc}} \right| &\leq \left( \frac{t}{t+m} \right)^{1-\gamma} - \frac{t}{(t+m+1)\left((t+1)/(t+m+1)\right)^\gamma}\notag\\
&\leq 1 - \bfrac{t+m}{t+m+1}^{1-\g}\bfrac{t}{t+1}^\g.\label{pr-th21-ts-4}
\end{align}
This follows since the right-hand sides of \eqref{pr-th21-ts-3} and \eqref{pr-th21-ts-4} are both monotonically decreasing in $t$.
Now 
\begin{align*}
A_{\aind}\leq &\frac{\left|\Wz\right|}{m(1+T)\lr{T/(T+1)}^\gamma} + \left[1-\bfrac{T+1}{T+1+1/m}^{1-\g}\bfrac{T}{T+1/m}^\g\right]\left|\Wz\right|\\
\leq &\left|\Wz\right|\left\{(1+T)^{\g-1}T^\g + \left[1-\bfrac{T+1}{T+2}^{1-\g}\bfrac{T}{T+1}^\g\right]\right\}.
\end{align*}
Therefore for almost every \oinO there exists a $T_3 = T_3(\omega)>0$ and independent of $m$ such that
\begin{equation*}
\sup_{mT_3\leq t< \infty}\supst\left|\frac{t-s}{u(t/m)}-\frac{\ceil{t}-\ceil{s}}{u(\tmc)}\right|\frac{\left|\Wz\right|}{m}<\frac{\eps}{2}.
\end{equation*}
Finally we have for almost every \oinO and with $T:=\max(T_1,T_2,T_3)$ that
\begin{align*}
&\left|\supmTk \Rpik - \supTt \Rpit \right|<\eps,
\end{align*}
since clearly $\sup_{mT\leq t<\infty} \Rpitm = \sup_{T\leq t<\infty} \Rpit$ for every $m$.
Putting these together we get
\begin{equation}
\supk \Rpik \limarm\supt \Rpit\as\notag
\end{equation}
and thus
\begin{equation*}
\supk \ofrac{\gtilde} \maxnik\abs{W_{1,m}(k)-W_{1,m}(i) - \frac{k-i}{m}W_{0,m}(m)}\stackrel{\mathcal{D}}{\longrightarrow}\supt\Rpit.
\end{equation*}
By computing the covariance functions it can be shown that
\begin{equation*}
\left\{\We{t}-t\Wz,\;0\leq t<\infty\right\}\eqD\left\{(1+t)W\left(t/(1+t)\right),\;0\leq t<\infty\right\}, 
\end{equation*}
where $\{W(t),\;0\leq t<\infty\}$ is again a Wiener process (cf. \citet{2004}). We conclude
\begin{align}
 \supt \Rpit\eqD&\supt \supst \frac{\abs{(1+t)\Wd{t/(1+t)} - (1+s)\Wd{s/(1+s)}}}{(1+t)\lr{t/(1+t)}^\gamma}\notag\\
 =&\supt \supst \frac{\abs{\Wd{t/(1+t)} - (({1+s})/({1+t}))\Wd{s/(1+s)}}}{\lr{t/(1+t)}^\gamma}\notag\\
 =&\sup_{0< t< 1} \supst \ofrac{t^\gamma}{\abs{W(t)-(({1-t})/({1-s}))W(s)}}.\notag
\end{align}
The proof can now be completed by taking the weak consistency of the estimator $\hat{\sigma}_m$ into account.
\prendwol

\subsection{Proof of Theorem \ref{Th2.2}}\label{P-Th2.2}
We only prove part a) of Theorem \ref{Th2.2}, parts b) and c) then follow immediately.
Since 
\begin{equation*}
\minnik\sumeh{j}{i}\leq 0,
\end{equation*}
we have 
\begin{equation*}
\Q=\sumeh{i}{k} \quad\leq \quad\sumeh{i}{k} -\minnik\sumeh{j}{i}
= \Po.
\end{equation*}
Consequently it is sufficient to show that under $H_A$ and $\bd^T\bDelta>0$ we have
\begin{equation}
\frac{1}{\hat{\sigma}_m}\sup_{1\leq k<\infty}\frac{\Q}{\gtilde} \limP \infty.\label{19}
\end{equation}
First note that, for $k\geq \kb$,

\begin{align}
 \Q = \sume ik &+ \lr{\sum_{i=m+1}^{m+k}\mbf{x}_i}^T(\bbeta_0 - \betam)\label{exp}\\
 &+ \lr{\sum_{i=m+\kb}^{m+k}(\mbf{x}_i - \bd)}^T\bDelta + (k-\kb+1)\bd^T\bDelta.\notag
\end{align}
From the proof of Theorem \ref{Th2.1-ts} we get
\begin{equation*}
 \supk \ofrac{\gtilde} \left|\sume ik + \lr{\sum_{i=m+1}^{m+k}\mbf{x}_i}^T(\bbeta_0 - \betam)\right| = \Op.
\end{equation*}
Now because of \eqref{L1-P2},
\begin{equation*}
 \lr{\sum_{i=m+\kb}^{m+k}(\mbf{x}_i - \bd)}^T\bDelta = o(k-\kb)\quad\text{as }k\to\infty,\qquad\text{a.s., uniformly in }m.
\end{equation*}
As a consequence the drift term $\supk(k-\kb+1)\bd^T\bDelta/\gtilde$ is the dominating term and it clearly diverges as $m\to\infty.$\prendwol
\textcolor{db}{
\subsection{Proofs of Theorems \ref{Th3.4} and \ref{Th3.6}}\label{P-Th3.4&3.6}
\citet{AHHK2006} showed similar results for the squared prediction errors. In case of a one-step prediction, the prediction errors are the recursive residuals, i.e. in the notation of \citet{AHHK2006}, $y_i - \hat{y}_i = y_i - \bx_i^T\hat{\bbeta}_{i-1} = \tilde{\eps}_i$.
These results hold as well when the recursive residuals are replaced by the ordinary residuals $\hat{\eps}_i$.
Therefore the proofs are a combination of the proofs in \citet{AHHK2006} with the arguments in the proofs of Theorems \ref{Th2.1-ts} and \ref{Th2.2}. It should be mentioned that the drift term under $H_A$ is determined by $\bDelta^T \bC\bDelta$ and therefore positive even under $\bd^T\bDelta = 0$.
}
\begin{appendix}
\refstepcounter{section}
\begin{table}[hptb]
\addcontentsline{toc}{section}{\numberline{A}Appendix} 
\begin{center}
\begin{tabular}{lcccccccccc} 
$\kb = 1$&\multicolumn{5}{c}{\bf $m =  200$}&\multicolumn{5}{c}{\bf $m =  1000$}\\
\cmidrule(r){1-6}\cmidrule(l){7-11}
\multicolumn{1}{l}{$\g = 0.00$}&\multicolumn{1}{c}{$\min$}&\multicolumn{1}{c}{$1^\text{st}$Q}&\multicolumn{1}{c}{med}&\multicolumn{1}{c}{$3^\text{rd}$Q}&\multicolumn{1}{c}{$\max$}&\multicolumn{1}{c}{$\min$}&\multicolumn{1}{c}{$1^\text{st}$Q}&\multicolumn{1}{c}{med}&\multicolumn{1}{c}{$3^\text{rd}$Q}&\multicolumn{1}{c}{$\max$}\\
\cmidrule(lr){1-6}\cmidrule(lr){7-11}
$\widehat{Q}_P$&~~15&~~39&~~48&~~60&~159
&~~48&~~84&~~96&~110&~191\\
$|\widehat{Q}|$&~~14&~~39&~~48&~~60&~159
&~~48&~~84&~~95&~109&~189\\ 
\cmidrule(lr){1-6}\cmidrule(lr){7-11}
$\widehat{Q}_P^u$&~~11&~~31&~~39&~~49&~142
&~~38&~~70&~~80&~~92&~179\\ 
$\widehat{Q}^u$&~~11&~~31&~~39&~~49&~143
&~~38&~~68&~~79&~~91&~177\\ 
\cmidrule(lr){1-6}\cmidrule(lr){7-11}
$\widehat{S}_P$&~~~3&~~32&~~46&~~68&~558
&~~22&~~76&~~96&~120&~256\\
$|\widehat{S}_R|$&~~~3&~~32&~~47&~~69&~558
&~~22&~~75&~~96&~120&~263\\ 
\cmidrule(lr){1-6}\cmidrule(lr){7-11}
$\widehat{S}^u_P$&~~~3&~~25&~~38&~~56&~449
&~~11&~~62&~~80&~102&~251\\ 
$\widehat{S}_R$&~~~3&~~25&~~38&~~57&~451
&~~11&~~61&~~80&~101&~251\\ 
\midrule
\multicolumn{1}{l}{$\g =  0.25$}&\multicolumn{1}{c}{$\min$}&\multicolumn{1}{c}{$1^\text{st}$Q}&\multicolumn{1}{c}{med}&\multicolumn{1}{c}{$3^\text{rd}$Q}&\multicolumn{1}{c}{$\max$}&\multicolumn{1}{c}{$\min$}&\multicolumn{1}{c}{$1^\text{st}$Q}&\multicolumn{1}{c}{med}&\multicolumn{1}{c}{$3^\text{rd}$Q}&\multicolumn{1}{c}{$\max$}
\\\cmidrule(lr){1-6}\cmidrule(lr){7-11}
$\widehat{Q}_P$&~~~4&~~20&~~28&~~38&~144
&~~12&~~35&~~44&~~56&~146\\ 
$|\widehat{Q}|$&~~~4&~~20&~~28&~~39&~144
&~~11&~~34&~~44&~~55&~148\\ 
\cmidrule(lr){1-6}\cmidrule(lr){7-11}
$\widehat{Q}_P^u$&~~~4&~~16&~~22&~~31&~108
&~~~9&~~28&~~37&~~47&~135\\ 
$\widehat{Q}^u$&~~~3&~~15&~~22&~~31&~107
&~~~8&~~27&~~36&~~46&~135\\ 
\cmidrule(lr){1-6}\cmidrule(lr){7-11}
$\widehat{S}_P$&~~~1&~~14&~~26&~~44&~558
&~~~1&~~28&~~44&~~64&~216\\ 
$|\widehat{S}_R|$&~~~1&~~15&~~27&~~46&~558
&~~~1&~~29&~~44&~~64&~217\\ 
\cmidrule(lr){1-6}\cmidrule(lr){7-11}
$\widehat{S}^u_P$&~~~1&~~11&~~21&~~37&~451
&~~~1&~~22&~~36&~~54&~202\\ 
$\widehat{S}_R$&~~~1&~~11&~~22&~~38&~451
&~~~1&~~22&~~35&~~53&~196\\ 
\midrule
\multicolumn{1}{l}{$\g =  0.49$}&\multicolumn{1}{c}{$\min$}&\multicolumn{1}{c}{$1^\text{st}$Q}&\multicolumn{1}{c}{med}&\multicolumn{1}{c}{$3^\text{rd}$Q}&\multicolumn{1}{c}{$\max$}&\multicolumn{1}{c}{$\min$}&\multicolumn{1}{c}{$1^\text{st}$Q}&\multicolumn{1}{c}{med}&\multicolumn{1}{c}{$3^\text{rd}$Q}&\multicolumn{1}{c}{$\max$}
\\ \cmidrule(lr){1-6}\cmidrule(lr){7-11}
$\widehat{Q}_P$&~~~1&~~~9&~~17&~~29&~148
&~~~1&~~~9&~~16&~~26&~135\\
$|\widehat{Q}|$&~~~1&~~~9&~~17&~~28&~148
&~~~1&~~~9&~~16&~~26&~133\\ 
\cmidrule(lr){1-6}\cmidrule(lr){7-11}
$\widehat{Q}_P^u$&~~~1&~~~8&~~14&~~24&~137
&~~~1&~~~8&~~13&~~22&~124\\ 
$\widehat{Q}^u$&~~~1&~~~7&~~14&~~24&~144
&~~~1&~~~7&~~13&~~22&~117\\ 
\cmidrule(lr){1-6}\cmidrule(lr){7-11}
$\widehat{S}_P$&~~~1&~~~6&~~15&~~33&1888
&~~~1&~~~6&~~14&~~30&~211\\ 
$|\widehat{S}_R|$&~~~1&~~~6&~~16&~~35&1460
&~~~1&~~~6&~~15&~~32&~200\\ 
\cmidrule(lr){1-6}\cmidrule(lr){7-11}
$\widehat{S}^u_P$&~~~1&~~~5&~~13&~~29&1269
&~~~1&~~~5&~~12&~~27&~197\\ 
$\widehat{S}_R$&~~~1&~~~5&~~13&~~30&~838
&~~~1&~~~5&~~13&~~28&~190\\ 
\bottomrule
\end{tabular}
\caption{Five number summary under $H_1$ with an early-change $\kb =1$ for $\alpha = 0.1$.}
\label{T6}
\end{center}
\end{table}

\begin{table}
\begin{center}
\begin{tabular}{lcccccccccc} 
$\kb = m$&\multicolumn{5}{c}{\bf $m =  200$}&\multicolumn{5}{c}{\bf $m =  1000$}\\
\cmidrule(r){1-6}\cmidrule(l){7-11}
\multicolumn{1}{l}{$\g = 0.00$}&\multicolumn{1}{c}{$\min$}&\multicolumn{1}{c}{$1^\text{st}$Q}&\multicolumn{1}{c}{med}&\multicolumn{1}{c}{$3^\text{rd}$Q}&\multicolumn{1}{c}{$\max$}&\multicolumn{1}{c}{$\min$}&\multicolumn{1}{c}{$1^\text{st}$Q}&\multicolumn{1}{c}{med}&\multicolumn{1}{c}{$3^\text{rd}$Q}&\multicolumn{1}{c}{$\max$}\\
\cmidrule(lr){1-6}\cmidrule(lr){7-11}
$\widehat{Q}_P$&~~~1&~~59&~~79&~101&~289
&~~~4&~126&~159&~189&~318\\
$|\widehat{Q}|$&~~~1&~~66&~~95&~131&~443
&~~~9&~143&~193&~248&~476\\ 
\cmidrule(lr){1-6}\cmidrule(lr){7-11}
$\widehat{Q}_P^u$&~~~1&~~45&~~63&~~82&~261
&~~~1&~~98&~129&~157&~263\\ 
$\widehat{Q}^u$&~~~2&~~51&~~78&~110&~450
&~~~1&~113&~160&~214&~558\\ 
\cmidrule(lr){1-6}\cmidrule(lr){7-11}
$\widehat{S}_P$&~~~1&~~46&~~75&~112&2200
&~~~2&~114&~156&~200&~467\\ 
$|\widehat{S}_R|$&~~~1&~~53&~~90&~141&2200
&~~~6&~130&~188&~258&~767\\ 
\cmidrule(lr){1-6}\cmidrule(lr){7-11}
$\widehat{S}^u_P$&~~~1&~~36&~~60&~~91&1593
&~~~1&~~90&~127&~165&~403\\ 
$\widehat{S}_R$&~~~1&~~41&~~74&~117&1595
&~~~1&~103&~157&~222&~650\\ 
\midrule
\multicolumn{1}{l}{$\g =  0.25$}&\multicolumn{1}{c}{$\min$}&\multicolumn{1}{c}{$1^\text{st}$Q}&\multicolumn{1}{c}{med}&\multicolumn{1}{c}{$3^\text{rd}$Q}&\multicolumn{1}{c}{$\max$}&\multicolumn{1}{c}{$\min$}&\multicolumn{1}{c}{$1^\text{st}$Q}&\multicolumn{1}{c}{med}&\multicolumn{1}{c}{$3^\text{rd}$Q}&\multicolumn{1}{c}{$\max$}
\\\cmidrule(lr){1-6}\cmidrule(lr){7-11}
$\widehat{Q}_P$&~~~2&~~54&~~73&~~96&~289
&~~~3&~113&~145&~175&~305\\ 
$|\widehat{Q}|$&~~~1&~~61&~~89&~124&~423
  &~~~2&~129&~177&~231&~451\\ 
\cmidrule(lr){1-6}\cmidrule(lr){7-11}
$\widehat{Q}_P^u$&~~~1&~~43&~~62&~~80&~290
&~~~1&~~93&~124&~150&~256\\ 
$\widehat{Q}^u$&~~~1&~~49&~~75&~108&~464
&~~~1&~107&~153&~206&~556\\ 
\cmidrule(lr){1-6}\cmidrule(lr){7-11}
$\widehat{S}_P$&~~~1&~~43&~~71&~107&2200
&~~~3&~104&~143&~186&~436\\ 
$|\widehat{S}_R|$&~~~1&~~49&~~85&~136&2200
&~~~3&~118&~175&~242&~731\\ 
\cmidrule(lr){1-6}\cmidrule(lr){7-11}
$\widehat{S}^u_P$&~~~1&~~36&~~60&~~91&1644
&~~~1&~~87&~122&~159&~380\\ 
$\widehat{S}_R$&~~~1&~~41&~~73&~117&1726
&~~~1&~~99&~151&~215&~648\\ 
\midrule
\multicolumn{1}{l}{$\g =  0.49$}&\multicolumn{1}{c}{$\min$}&\multicolumn{1}{c}{$1^\text{st}$Q}&\multicolumn{1}{c}{med}&\multicolumn{1}{c}{$3^\text{rd}$Q}&\multicolumn{1}{c}{$\max$}&\multicolumn{1}{c}{$\min$}&\multicolumn{1}{c}{$1^\text{st}$Q}&\multicolumn{1}{c}{med}&\multicolumn{1}{c}{$3^\text{rd}$Q}&\multicolumn{1}{c}{$\max$}
\\ \cmidrule(lr){1-6}\cmidrule(lr){7-11}
$\widehat{Q}_P$&~~~3&~~77&~103&~133&~490
&~~14&~153&~189&~223&~386\\ 
$|\widehat{Q}|$&~~~2&~~81&~115&~160&~658
&~~15&~162&~214&~273&~526\\ 
\cmidrule(lr){1-6}\cmidrule(lr){7-11}
$\widehat{Q}_P^u$&~~~2&~~67&~~91&~119&~430
&~~~6&~136&~170&~202&~342\\ 
$\widehat{Q}^u$&~~~1&~~71&~103&~144&~569
&~~~7&~144&~196&~254&~631\\ 
\cmidrule(lr){1-6}\cmidrule(lr){7-11}
$\widehat{S}_P$&~~~2&~~63&~100&~155&2200
&~~~1&~140&~187&~237&~641\\ 
$|\widehat{S}_R|$&~~~1&~~67&~111&~181&2200
&~~~5&~151&~212&~288&~890\\ 
\cmidrule(lr){1-6}\cmidrule(lr){7-11}
$\widehat{S}^u_P$&~~~1&~~55&~~90&~138&2200
&~~~2&~125&~169&~217&~522\\ 
$\widehat{S}_R$&~~~1&~~59&~100&~163&2200
&~~~3&~135&~194&~267&~826\\ 
\bottomrule
\end{tabular}
\caption{Five number summary under $H_1$ with $\kb =m$ for $\alpha = 0.1$
.}
\label{T7}
\end{center}
\end{table}
\begin{table}
\begin{center}
\begin{tabular}{lcccccccccc} 
$\kb = 5m$&\multicolumn{5}{c}{\bf $m =  200$}&\multicolumn{5}{c}{\bf $m =  1000$}\\
\cmidrule(r){1-6}\cmidrule(l){7-11}
\multicolumn{1}{l}{$\g = 0.00$}&\multicolumn{1}{c}{$\min$}&\multicolumn{1}{c}{$1^\text{st}$Q}&\multicolumn{1}{c}{med}&\multicolumn{1}{c}{$3^\text{rd}$Q}&\multicolumn{1}{c}{$\max$}&\multicolumn{1}{c}{$\min$}&\multicolumn{1}{c}{$1^\text{st}$Q}&\multicolumn{1}{c}{med}&\multicolumn{1}{c}{$3^\text{rd}$Q}&\multicolumn{1}{c}{$\max$}\\
\cmidrule(lr){1-6}\cmidrule(lr){7-11}
$\widehat{Q}_P$&~~~2&~182&~256&~313&~550
&~~11&~382&~506&~581&~785\\ 
$|\widehat{Q}|$&~~~2&~191&~293&~412&~933
&~~~6&~404&~577&~777&1436\\ 
\cmidrule(lr){1-6}\cmidrule(lr){7-11}
$\widehat{Q}_P^u$&~~~3&~152&~212&~259&~489
&~~~5&~319&~425&~490&~666\\ 
$\widehat{Q}^u$&~~~2&~158&~252&~366&1199
&~~~2&~340&~503&~700&1826\\ 
\cmidrule(lr){1-6}\cmidrule(lr){7-11}
$\widehat{S}_P$&~~~2&~150&~234&~331&3000
&~~~6&~352&~492&~601&1074\\ 
$|\widehat{S}_R|$&~~~1&~158&~263&~418&3000
&~~~1&~369&~566&~777&1817\\ 
\cmidrule(lr){1-6}\cmidrule(lr){7-11}
$\widehat{S}^u_P$&~~~1&~126&~198&~278&1634
&~~13&~296&~414&~511&1027\\ 
$\widehat{S}_R$&~~~1&~133&~225&~366&3000
&~~~5&~311&~491&~696&1938\\ 
\midrule
\multicolumn{1}{l}{$\g =  0.25$}&\multicolumn{1}{c}{$\min$}&\multicolumn{1}{c}{$1^\text{st}$Q}&\multicolumn{1}{c}{med}&\multicolumn{1}{c}{$3^\text{rd}$Q}&\multicolumn{1}{c}{$\max$}&\multicolumn{1}{c}{$\min$}&\multicolumn{1}{c}{$1^\text{st}$Q}&\multicolumn{1}{c}{med}&\multicolumn{1}{c}{$3^\text{rd}$Q}&\multicolumn{1}{c}{$\max$}
\\\cmidrule(lr){1-6}\cmidrule(lr){7-11}
$\widehat{Q}_P$&~~~7&~199&~276&~335&~608
&~~~8&~411&~537&~613&~826\\ 
$|\widehat{Q}|$&~~~3&~205&~310&~432&1035
&~~~8&~427&~601&~801&1471\\ 
\cmidrule(lr){1-6}\cmidrule(lr){7-11}
$\widehat{Q}_P^u$&~~~2&~172&~236&~288&~604
&~~~2&~357&~466&~533&~713\\ 
$\widehat{Q}^u$&~~~2&~176&~273&~391&1248
&~~~9&~372&~536&~737&1857\\ 
\cmidrule(lr){1-6}\cmidrule(lr){7-11}
$\widehat{S}_P$&~~~3&~166&~255&~361&3000
&~~11&~384&~523&~638&1228\\ 
$|\widehat{S}_R|$&~~~1&~172&~280&~443&3000
&~~15&~395&~593&~808&1895\\ 
\cmidrule(lr){1-6}\cmidrule(lr){7-11}
$\widehat{S}^u_P$&~~~2&~142&~221&~311&3000
&~~15&~333&~456&~555&1123\\ 
$\widehat{S}_R$&~~~3&~149&~247&~395&3000
&~~~3&~343&~527&~733&1977\\ 
\midrule
\multicolumn{1}{l}{$\g =  0.49$}&\multicolumn{1}{c}{$\min$}&\multicolumn{1}{c}{$1^\text{st}$Q}&\multicolumn{1}{c}{med}&\multicolumn{1}{c}{$3^\text{rd}$Q}&\multicolumn{1}{c}{$\max$}&\multicolumn{1}{c}{$\min$}&\multicolumn{1}{c}{$1^\text{st}$Q}&\multicolumn{1}{c}{med}&\multicolumn{1}{c}{$3^\text{rd}$Q}&\multicolumn{1}{c}{$\max$}
\\ \cmidrule(lr){1-6}\cmidrule(lr){7-11}
$\widehat{Q}_P$&~~~1&~326&~430&~526&1089
&~~34&~638&~785&~876&1203\\ 
$|\widehat{Q}|$&~~23&~315&~450&~610&1577
&~~17&~622&~819&1049&1898\\ 
\cmidrule(lr){1-6}\cmidrule(lr){7-11}
$\widehat{Q}_P^u$&~~12&~292&~386&~469&~977
&~~30&~578&~715&~802&1085\\ 
$\widehat{Q}^u$&~~13&~278&~402&~553&1679
&~~19&~558&~746&~975&7000\\ 
\cmidrule(lr){1-6}\cmidrule(lr){7-11}
$\widehat{S}_P$&~~11&~268&~399&~574&3000
&~~36&~595&~766&~912&1584\\ 
$|\widehat{S}_R|$&~~~3&~256&~407&~641&3000
&~~22&~580&~808&1059&7000\\ 
\cmidrule(lr){1-6}\cmidrule(lr){7-11}
$\widehat{S}^u_P$&~~~1&~239&~356&~508&3000
&~~~6&~537&~700&~833&1468\\ 
$\widehat{S}_R$&~~~6&~226&~363&~575&3000
&~~~3&~516&~736&~977&7000\\ 
\bottomrule
\end{tabular}
\caption{Five number summary under $H_1$ with $\kb =5m$ for $\alpha = 0.1$
.}
\label{T8}
\end{center}
\end{table}

\begin{table}
\begin{center}
\begin{tabular}{llccccc
} 
\toprule
&\multicolumn{1}{l}{}&\multicolumn{1}{c}{$\min$}&\multicolumn{1}{c}{$1^\text{st}$Q}&\multicolumn{1}{c}{med}&\multicolumn{1}{c}{$3^\text{rd}$Q}&\multicolumn{1}{c}{$\max$}
\\
\cmidrule(lr){1-7}
$\kb = 1$&$\widehat{S}_P$&~~10&~~58&~~75&~~94&~235\\ 
&$|\widehat{S}_R|$&~~10&~~58&~~75&~~94&~235\\ 
\cmidrule(lr){2-7}
&$\widehat{S}^u_P$&~~~8&~~48&~~63&~~80&~178\\ 
&$\widehat{S}_R$&~~~8&~~47&~~62&~~79&~178\\ 
\cmidrule(lr){1-7}
$\kb = m$&$\widehat{S}_P$&~~~2&~~86&~119&~155&~353\\ 
&$|\widehat{S}_R|$&~~~2&~~98&~144&~196&~537\\ 
\cmidrule(lr){2-7}
&$\widehat{S}^u_P$&~~~1&~~68&~~98&~129&~293\\ 
&$\widehat{S}_R$&~~~1&~~78&~121&~169&~493\\ 
\cmidrule(lr){1-7}
$\kb = 5m$&$\widehat{S}_P$&~~~4&~271&~376&~461&~823\\ 
&$|\widehat{S}_R|$&~~~4&~287&~435&~590&1345\\ 
\cmidrule(lr){2-7}
&$\widehat{S}^u_P$&~~~7&~228&~317&~392&~676\\ 
&$\widehat{S}_R$&~~~6&~243&~380&~532&1526\\ 
\bottomrule
\end{tabular}
\caption{Five number summary under $H_2$ for $\alpha = 0.1$, $\g = 0.00$ and $m = 1000$.}
\label{T9}
\end{center}
\end{table}
\end{appendix}

\bibliographystyle{plainnat}

\begin{thebibliography}{28}
\providecommand{\natexlab}[1]{#1}
\providecommand{\url}[1]{\texttt{#1}}
\expandafter\ifx\csname urlstyle\endcsname\relax
  \providecommand{\doi}[1]{doi: #1}\else
  \providecommand{\doi}{doi: \begingroup \urlstyle{rm}\Url}\fi

\bibitem[Antoch and Jaru\v{s}kov{\'a}(2002)]{antoch2002}
J.~Antoch and D.~Jaru\v{s}kov{\'a}.
\newblock On-line statistical process control.
\newblock In \emph{Multivariate Total Quality Control, Foundations and Recent
  Advances. Eds: Lauro, C., Antoch, J. and Vinzi, V.E.} Physica, Heidelberg,
  2002.

  \bibitem[Aue et~al.(2006{\natexlab{a}})Aue, Berkes and Horv{\'a}th]{ABH2006}
A.~Aue, I.~Berkes and L.~Horv{\'a}th.
\newblock {Strong approximation for the sums of squares of augmented GARCH
  sequences}.
\newblock \emph{Bernoulli}, 12\penalty0 (4):\penalty0 583, 2006{\natexlab{a}}.

  \bibitem[Aue et~al.(2013)Aue, Dienes, Fremdt and Steinebach]{ADFS2013}
A.~Aue, C.~Dienes, S.~Fremdt, J.G.~Steinebach.
\newblock {Reaction times of monitoring schemes for ARMA time series}.
\newblock \emph{Preprint University of California Davis, University of Cologne}. 2013.

\bibitem[Aue and Horv{\'a}th(2004)]{AH2004}
A.~Aue and L.~Horv{\'a}th.
\newblock {Delay time in sequential detection of change}.
\newblock \emph{Statistics \& Probability Letters}, 67\penalty0 (3):\penalty0
  221--231, 2004.

\bibitem[Aue and Horv{\'a}th(2013)]{AH2013}
A.~Aue and L.~Horv{\'a}th.
\newblock {Structural breaks in time series}.
\newblock \emph{Journal of Time Series Analysis}, 34\penalty0 (1):\penalty0
  1--16, 2013.
  
  
\bibitem[Aue et~al.(2006{\natexlab{b}})Aue, Horv{\'a}th, Hu\v{s}kov{\'a} and
  Kokoszka]{AHHK2006}
A.~Aue, L.~Horv{\'a}th, M.~Hu\v{s}kov{\'a} and P.~Kokoszka.
\newblock {Change-point monitoring in linear models}.
\newblock \emph{Econometrics Journal}, 9\penalty0 (3):\penalty0 373--403,
  2006{\natexlab{b}}.

\bibitem[Aue et~al.(2012)Aue, Horv{\'a}th, K{\"u}hn and
  Steinebach]{aue2008reaction}
A.~Aue, L.~Horv{\'a}th, M.~K{\"u}hn and J.~Steinebach.
\newblock {On the reaction time of moving sum detectors}.
\newblock \emph{Journal of Statistical Planning and Inference}, 142\penalty0
  (8):\penalty0 2271--2288, 2012.

\bibitem[Aue et~al.(2009)Aue, Horv{\'a}th and Reimherr]{AHR2007}
A.~Aue, L.~Horv{\'a}th and M.L. Reimherr.
\newblock Delay times of sequential procedures for multiple time series
  regression models.
\newblock \emph{Journal of Econometrics}, 149\penalty0 (2):\penalty0 174 --
  190, 2009.

\bibitem[Aue et~al.(2012)Aue, H{\"o}rmann, Horv{\'a}th, Hu\v{s}kov{\'a} and
  Steinebach]{AHHHS2010}
A.~Aue, S.~H{\"o}rmann, L.~Horv{\'a}th, M.~Hu\v{s}kov{\'a} and J.G.
  Steinebach.
\newblock Sequential testing for the stability of high frequency portfolio
  betas.
\newblock \emph{Econometric Theory}, 28\penalty0 (4):\penalty0 804-837, 2012.

\bibitem[Bai(1997)]{bai1997}
J.~Bai.
\newblock Estimation of a change point in multiple regression models.
\newblock \emph{Review of Economics and Statistics}, 79\penalty0 (4):\penalty0
  551--563, 1997.

\bibitem[Carrasco and Chen(2002)]{carrasco2002}
M.~Carrasco and X.~Chen.
\newblock Mixing and moment properties of various {GARCH} and stochastic
  volatility models.
\newblock \emph{Econometric Theory}, 18\penalty0 (1):\penalty0 17--39, 2002.

\bibitem[Chu et~al.(1995)Chu, Hornik and Kuan]{chu1995mosum}
C.S.J. Chu, K.~Hornik and C.M. Kuan.
\newblock MOSUM tests for parameter constancy.
\newblock \emph{Biometrika}, 82\penalty0 (3):\penalty0 603, 1995.

\bibitem[Chu et~al.(1996)Chu, Stinchcombe and White]{1996}
C.S.J. Chu, M.~Stinchcombe and H.~White.
\newblock Monitoring structural change.
\newblock \emph{Econometrica}, 64\penalty0 (5):\penalty0 1045--1065, 1996.

\bibitem[Cs{\"o}rg\H{o} and Horv{\'a}th(1997)]{CH1997}
M.~Cs{\"o}rg\H{o} and L.~Horv{\'a}th.
\newblock \emph{Limit Theorems in Change-Point Analysis.}
\newblock Wiley, Chichester, 1997.

\bibitem[Cs{\"o}rg\H{o} and R{\'e}v{\'e}sz(1981)]{1981}
M.~Cs{\"o}rg\H{o} and P.~R{\'e}v{\'e}sz.
\newblock \emph{Strong Approximations in Probability and Statistics}.
\newblock Academic Press, 1981.

\bibitem[Duan(1997)]{Duan1997}
J.C. Duan.
\newblock Augmented {GARCH} (p,q) process and its diffusion limit.
\newblock \emph{Journal of Econometrics}, 79\penalty0 (1):\penalty0 97--127,
  1997.

\bibitem[Fama and French(1993)]{Fama1993}
E.F. Fama and K.R. French.
\newblock Common risk factors in the returns on stocks and bonds.
\newblock \emph{Journal of Financial Economics}, 33\penalty0 (1):\penalty0
  3--56, 1993.

\bibitem[Fama and French(1996)]{FF1996}
E.F. Fama and K.R. French.
\newblock Multifactor explanations of asset pricing anomalies.
\newblock \emph{Journal of Finance}, 51\penalty0 (1):\penalty0 55--84, 1996.

\bibitem[French(2011)]{FF-website}
K.R. French.
\newblock Kenneth {R}. {F}rench - {D}ata {L}ibrary, July 2011.\\
\newblock URL
  \url{http://mba.tuck.dartmouth.edu/pages/faculty/ken.french/data\_library.ht%
ml}.

\bibitem[Garcia and Ghysels(1998)]{GG1998}
R.~Garcia and E.~Ghysels.
\newblock Structural change and asset pricing in emerging markets.
\newblock \emph{Journal of International Money and Finance}, 17\penalty0
  (3):\penalty0 455--473, 1998.

\bibitem[Horv{\'a}th et~al.(2004)Horv{\'a}th, Hu\v{s}kov{\'a}, Kokoszka and
  Steinebach]{2004}
L.~Horv{\'a}th, M.~Hu\v{s}kov{\'a}, P.~Kokoszka and J.~Steinebach.
\newblock Monitoring changes in linear models.
\newblock \emph{Journal of Statistical Planning and Inference}, 126\penalty0
  (1):\penalty0 225--251, 2004.

\bibitem[Hu\v{s}kov{\'a} and Koubkov{\'a}(2005)]{huskova2005}
M.~Hu\v{s}kov{\'a} and A.~Koubkov{\'a}.
\newblock Monitoring jump changes in linear models.
\newblock \emph{Journal of Statistical Research}, 39\penalty0 (2):\penalty0
  51--70, 2005.

\bibitem[Hu\v{s}kov{\'a} et~al.(2007)Hu\v{s}kov{\'a}, Pr{\'a}\v{s}kov{\'a} and
  Steinebach]{HPS2007}
M.~Hu\v{s}kov{\'a}, Z.~Pr{\'a}\v{s}kov{\'a} and J.~Steinebach.
\newblock On the detection of changes in autoregressive time series {I.
  A}symptotics.
\newblock \emph{Journal of {S}tatistical {P}lanning and {I}nference}, 137\penalty0
  (4):\penalty0 1243--1259, 2007.

\bibitem[Kothari et~al.(1995)Kothari, Shanken and Sloan]{KSS1995}
S.P. Kothari, J.~Shanken and R.G. Sloan.
\newblock Another look at the cross-section of expected stock returns.
\newblock \emph{Journal of {F}inance}, 50\penalty0 (1):\penalty0 185--224, 1995.

\bibitem[Lintner(1965)]{lintner1965valuation}
J.~Lintner.
\newblock The valuation of risk assets and the selection of risky investments
  in stock portfolios and capital budgets.
\newblock \emph{The Review of Economics and Statistics}, 47\penalty0
  (1):\penalty0 13--37, 1965.

\bibitem[MacKinlay(1995)]{MacKinlay1995}
A.C.~MacKinlay.
\newblock Multifactor models do not explain deviations from the {CAPM}.
\newblock \emph{Journal of Financial Economics}, 38\penalty0 (1):\penalty0
  3--28, 1995.

\bibitem[Page(1954)]{1954}
E.~S. Page.
\newblock Continuous inspection schemes.
\newblock \emph{Biometrika}, 41\penalty0 (1/2):\penalty0 100--115, 1954.

\bibitem[Perron(2006)]{perron2006}
P.~Perron.
\newblock Dealing with structural breaks.
\newblock \emph{Palgrave Handbook of Econometrics}, 1:\penalty0 278--352, 2006.

\bibitem[Schmitz and Steinebach(2010)]{SS2010}
A.~Schmitz and J.G. Steinebach.
\newblock A note on the monitoring of changes in linear models with dependent
  errors.
\newblock \emph{Dependence in Probability and Statistics, Lecture Notes in
  Statistics}, 200:\penalty0 159--174, 2010.

\bibitem[Sharpe(1964)]{Sharpe1964}
W.F. Sharpe.
\newblock Capital asset prices: {A} theory of market equilibrium under
  conditions of risk.
\newblock \emph{Journal of Finance}, 19\penalty0 (3):\penalty0 425--442, 1964.

\end{thebibliography}

\end{document}